\documentclass[%
  aps, pre, reprint,
  superscriptaddress, altaffillsymbol,
  floats, floatfix, showkeys,
  footinbib, longbibliography, amsmath
]{revtex4-2}

\usepackage{font-preamble}

\usepackage{xcolor}
\usepackage{graphicx}
\graphicspath{{./}}
\usepackage[%
  bookmarks,bookmarksopen=true,bookmarksopenlevel=2,%
  colorlinks=true,allcolors=blue]{hyperref}

\usepackage[capitalize]{cleveref}
\crefname{section}{Sec.}{Secs.}
\crefname{figure}{Fig.}{Figs.}

\makeatletter
\newcommand\setcurrentname[1]{\def\@currentlabelname{#1}}
\makeatother

\begin{document}

\title{%
  Nonadiabatic force matching
  for alchemical free-energy estimation}

\author{Jorge L.\ Rosa-Raíces}
\affiliation{%
  Department of Chemistry,
  University of California,
  Berkeley, California 94720, USA}

\author{David T.\ Limmer}
\email{dlimmer@berkeley.edu}
\affiliation{%
  Department of Chemistry,
  University of California,
  Berkeley, California 94720, USA}
\affiliation{%
  Materials Science Division,
  Lawrence Berkeley National Laboratory,
  Berkeley, California 94720, USA}
\affiliation{%
  Chemical Science Division,
  Lawrence Berkeley National Laboratory,
  Berkeley, California 94720, USA}
\affiliation{%
  Kavli Energy NanoScience Institute,
  Berkeley, California 94720, USA}

\begin{abstract}
We propose a method to compute free-energy differences from nonadiabatic alchemical transformations using flow-based generative models.
The method, nonadiabatic force matching, hinges on estimating the dissipation along an alchemical switching process in terms of a nonadiabatic force field that can be learned through stochastic flow matching.
The learned field can be used in conjunction with short-time trajectory data to evaluate upper and lower bounds on the alchemical free energy that variationally converge to the exact value if the field is optimal.
Applying the method to evaluate the alchemical free energy of atomistic models shows that it can substantially reduce the simulation cost of a free-energy estimate at negligible loss of accuracy when compared with thermodynamic integration.
\end{abstract}


\maketitle

\section*{Introduction}

Computational schemes for accurate, high-throughput free-energy estimation are highly coveted and could enable scientific breakthroughs in materials design~\cite{Alberi2019MaterialsDesign} and drug discovery~\cite{Sadybekov2023DrugDiscovery}.
Presently, widespread methods adopt an alchemical approach, in which the state of a system is transformed along a thermodynamic path that connects the two equilibrium states, \( \mathsf{A} \) and \( \mathsf{B} \), between which the free-energy difference \( \Delta F \) is sought~\cite{York2023ModernAlchemical}.
Changes in the system's state are driven by a switching potential \( U(\bm{x}, \lambda) \), where the switch \( \lambda \) runs from \( \lambda_\mathsf{A} \) to \( \lambda_\mathsf{B} \) and indexes a family of equilibrium probability densities
\begin{equation}\label{eq:EquilibriumDensity}
  \pi(\bm{x}, \lambda)
  \equiv \exp\{ \beta [F(\lambda) - U(\bm{x}, \lambda)] \}
\end{equation}
over system configurations \( \bm{x} \), with \( \beta^{-1} \) the temperature times Boltzmann's constant and \( F(\lambda) \) the absolute Helmholtz free energy at \( \lambda \).
If the system is driven at a slow enough rate to prevent dissipation throughout the switching, the free-energy difference between its end states is~\cite{Kirkwood1935StatisticalMechanics}
\begin{equation}\label{eq:TI}
  \Delta F
  \equiv F(\lambda_\mathsf{B}) - F(\lambda_\mathsf{A})
  = \int_{\lambda_\mathsf{A}}^{\lambda_\mathsf{B}}\negthinspace \mathrm{d}\lambda \thinspace \langle \partial_\lambda U \rangle_\lambda
\end{equation}
where the angled brackets \( \langle \,\cdot\, \rangle_\lambda \) denote a configurational average under the probability density in \cref{eq:EquilibriumDensity}.
The duration of this adiabatic alchemical transformation approaches the quasistatic limit, exceeding the system's longest relaxation timescales and allowing it to reach equilibrium at every \( \lambda \).

Free-energy estimation by numerical evaluation of \cref{eq:TI}, known as thermodynamic integration (TI), underpins the computation of solubility curves and binding affinities for molecular systems of modest size~\cite{Gobbo2019ComputingSolubility,Cournia2017RelativeBinding}.
However, TI entails sampling a locus of high-dimensional, multi-modal equilibrium states along the thermodynamic path~\cite{Mey2020BestPractices}, and is seldom tractable at the scale of modern campaigns for drug discovery and materials design.
An alternative avenue for free-energy estimation is suggested by Jarzynski's work theorem~\cite{Jarzynski1997NonequilibriumEquality}
\begin{equation}\label{eq:Jarzynski}
  \exp(-\beta \Delta F)
  = \langle \exp(-\beta \mathcal{W}_{\mspace{-2mu}\tau}) \rangle
\end{equation}
which relates \( \Delta F \) to the work \( \mathcal{W}_{\mspace{-2mu}\tau} \) done on the system to drive it through the thermodynamic path via a switching process of duration \( \tau \).
The expression in angled brackets \( \langle \,\cdot\, \rangle \) in \cref{eq:Jarzynski} denotes an average over nonequilibrium work measurements; its evaluation enables free-energy estimation from nonadiabatic alchemical transformations, which are those whose duration is short compared to the quasistatic limit.
Seeking computational expediency beyond that of TI without sacrificing accuracy, a flurry of schemes have emerged to evaluate \cref{eq:Jarzynski} with maximum efficiency and minimum bias~\cite{Oberhofer2005BiasedSampling,Shirts2007PrimeTime,Dellago2014Computing,Wan2023Comparison}.

Naive evaluation of Jarzynski's equality in \cref{eq:Jarzynski} is far from efficient, as converging the exponential average on its right-hand side often requires exponentially many work measurements~\cite{Jarzynski2006RareEvents}.
Instead, modern schemes for free-energy estimation based on Jarzynski's equality target the exponentially rare work measurements that dominate the average, deploying probabilistic generative modeling techniques~\cite{Ho2020DenoisingDiffusion,Papamakarios2021NormalizingFlows,Albergo2023StochasticInterpolants,Vargas2024Transport} to mine quasi-adiabatic paths connecting states \( \mathsf{A} \) and \( \mathsf{B} \)~\cite{Noe2019BoltzmannGenerators,Wirnsberger2020TargetedFE,Zhao2023BoundingFE,Zhong2024TimeAsymmetric,Mate2024NeuralTI,He2025FEAT}.
To address scenarios where such paths are intractable due to computational limitations, it is desirable to extend generative modeling approaches to free-energy estimation that leverage typical nonequilibrium work measurements from simulation or experiment~\cite{Liu2014DirectMeasurement,Ciliberto2017Experiments,Ross2018EquilibriumFE,Bechhoefer2020StochasticThermodynamics,Barker2022ExperimentalVerification}.
Necessary to this end are theoretical perspectives that clarify the dissipative contributions to the exponential average in \cref{eq:Jarzynski} in terms of quantities that can be estimated from readily accessible trajectory data.

In this article, we examine flow matching models (FMMs) as an emergent platform for dissipative free-energy estimation.
After formulating an alchemical interpretation of the FMM archetype and laying out the stochastic thermodynamics of the nonadiabatic transformation that it induces, we elucidate descriptors of the dissipation incurred along the transformation that yield sharp bounds on the free-energy difference.
The bounds are saturated by a flow field that extends a stationary analogue introduced in Ref.~\onlinecite{Rosa-Raices2024VariationalTimeReversal} to evaluate log-likelihood functions for active systems, and that can be learned by training a FMM.
Numerical demonstrations on atomistic models of fast-growth solvation and of Frenkel--Ladd solid formation confirm that, given a nonadiabatic alchemical transformation and an associated FMM, our variational bounds can produce a free-energy estimate with comparable accuracy to TI at a substantially reduced simulation cost.

\section*{The stochastic thermodynamics \\ of flow matching models}

\begin{figure}[ht]
\centering
\includegraphics[width=2.8in]%
  {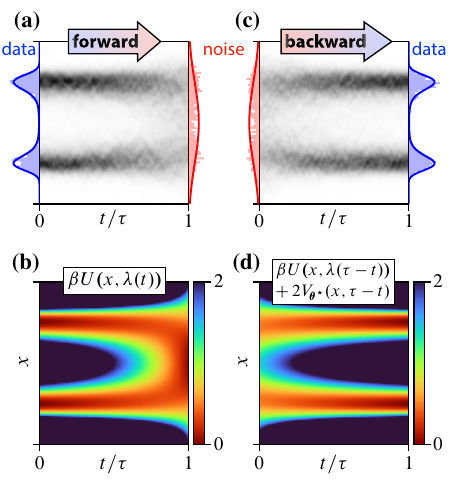}
\caption{%
  A flow matching model for nonadiabatic alchemical free-energy estimation
  \textbf{(a)}
  Sample realizations of the forward process that drives a system from state \( \mathsf{A} \), a bimodal distribution, toward state \(\mathsf{B} \), a Gaussian distribution, according to a switch \( \lambda(t) \) of duration \( \tau \).
  \textbf{(b)}
  The switching potential \( \beta U \bl(\bm{x}, \lambda(t)\br) \) drives the system away from a bimodal distribution and toward a Gaussian distribution.
  \textbf{(c)}
  The backward process, using the learned nonadiabatic potential \( V_{\!\bm{\theta}^\star}(\bm{x}, t) \), stochastically reverses the transition driven by the switching potential.
  \textbf{(d)}
  The nonadiabatic switching potential \( \beta U \bl(\bm{x}, \lambda(\tau\mspace{-1mu}\smallminus\mspace{-1mu}t) \br) + 2V_{\!\bm{\theta}^\star}(\bm{x}, \tau\mspace{-1mu}\smallminus\mspace{-1mu}t) \) restores the system to a bimodal distribution.
}\label{fig:nfm_schematic}
\end{figure}

FMMs learn a two-sided mapping between distributions of \emph{data} and \emph{noise} that allows generating samples from either distribution by transporting samples from the other.
The mapping is comprised of a pair of stochastic processes, of which the \emph{forward} process is given and the \emph{backward} process is its learned time-reversal~\cite{Song2021ScoreBased}.
Here we specify an alchemical FMM paradigm in which the map is determined by a switch \( \lambda(t) \) of duration \( \tau \) with the boundary values \( \lambda(t \smallequals 0) \equiv \lambda_\mathsf{A} \) and \( \lambda(t \smallequals \tau) \equiv \lambda_\mathsf{B} \), and where the forward process
\begin{equation}\label{eq:ForwardProcess}
  \mathrm{d}\bm{x}_{\mspace{-1mu}t} = -\beta D \bm\nabla U \bl( \bm{x}_{\mspace{-1mu}t}, \lambda(t) \br) \thinspace \mathrm{d}t + \sqrt{2D} \thinspace \mathrm{d}\bm{w}_{\mspace{-1mu}t}
\end{equation}
has an initial configuration \( \bm{x}_{\mspace{-1mu}0} \) drawn from the data distribution with density \( \pi(\bm{x}, \lambda_\mathsf{A}) \).
Here and throughout, \( D \) denotes a diffusion coefficient, \( \{ \bm{w}_{\mspace{-1mu}t} \}_{t=\mspace{1mu}0}^\tau \) a standard Brownian motion, and \( U \bl( \bm{x}_{\mspace{-1mu}t}, \lambda(t) \br) \) the switching potential introduced in \cref{eq:EquilibriumDensity}.
The forward process in \cref{eq:ForwardProcess} maps the data distribution onto the noise distribution with density \( \rho(\bm{x}, \tau) \thinspace\mathord{\equiv}\thinspace \langle \delta(\bm{x}_{\mspace{-1mu}\tau} \thinspace\mathord{-}\thinspace \bm{x}) \rangle \), where \( \delta \) is the Dirac delta and where the angled brackets \( \langle \,\cdot\, \rangle \) denote an average over realizations of the forward process.
The noise density satisfies the Fokker--Planck equation
\begin{equation}\label{eq:FokkerPlanck}
  \partial_t \rho(\bm{x}, t) = -\bm\nabla \cdot [ D \bm\nabla V(\bm{x}, t) \rho(\bm{x}, t) ]
\end{equation}
subject to the initial condition \( \rho(\bm{x}, t \smallequals 0) \equiv \pi(\bm{x}, \lambda_\mathsf{A}) \), and the nonadiabatic potential
\begin{equation}\label{eq:NonadiabaticPotential}
  V(\bm{x}, t)
  \equiv -\ln \frac{\rho(\bm{x}, t)}{\pi \bl( \bm{x}, \lambda(t) \br)}
\end{equation}
advects probability density from state \( \mathsf{A} \) toward state \( \mathsf{B} \).
Figure~\ref{fig:nfm_schematic} illustrates the alchemical FMM specified herein.

Physically, the forward process in \cref{eq:ForwardProcess} corresponds to a nonadiabatic alchemical transformation where a system, initially at thermal equilibrium in state \( \mathsf{A} \), is mechanically driven toward state \( \mathsf{B} \) in an irreversible manner.
Although this transformation occurs far from equilibrium, the free-energy difference \( \Delta F \) remains well-defined within the framework of stochastic thermodynamics~\cite{Limmer2024Book,Seifert2025Book}.
By the first law, the change of energy along the transformation is
\begin{equation}\label{eq:Energy}
  \Delta U
  \equiv \langle U \bl( \bm{x}_{\mspace{-1mu}\tau}, \lambda(\tau) \br) - U \bl( \bm{x}_{\mspace{-1mu}0}, \lambda(0) \br) \rangle
  = \langle \mathcal{W}_{\mspace{-2mu}\tau} \rangle - \langle \mathcal{Q}_\tau \rangle
\end{equation}
with
\begin{subequations}\label{eq:ForwardPathFunctionals}
\begin{align}
  \mathcal{W}_{\mspace{-2mu}\tau}
  &\equiv \int_0^{\mspace{-1mu}\tau}\negthickspace \mathrm{d}t \thinspace \partial_t U \bl( \bm{x}_{\mspace{-1mu}t}, \lambda(t) \br)
  \label{eq:PathWork}
  \\ \mathcal{Q}_\tau &\equiv -\int_0^{\mspace{-1mu}\tau}\negthickspace \mathrm{d}\bm{x}_{\mspace{-1mu}t} \mspace{2mu}\mathord{\circ}\mspace{2mu} \bm\nabla U \bl( \bm{x}_{\mspace{-1mu}t}, \lambda(t) \br) \label{eq:PathHeat}
\end{align}
\end{subequations}
the work exerted on the system and the heat dissipated by the system, respectively, along a realization of the process.
Using \cref{eq:EquilibriumDensity}, the first equality in \cref{eq:Energy} reduces to an expression for the Helmholtz free-energy difference, namely (\nameref{sec:SI}, Section~I)
\begin{equation}\label{eq:FreeEnergy}
  \Delta F = \Delta U - \beta^{-1} [ \Delta S + D_\mathrm{KL}(\tau) ]
\end{equation}
where
\( \Delta S \equiv \langle -\ln[ \rho(\bm{x}_{\mspace{-1mu}\tau}, \tau) / \pi(\bm{x}_{\mspace{-1mu}0}, \lambda_\mathsf{A}) ] \rangle \)
is the entropy change of the system along the transformation; the Kullback--Leibler (KL) divergence
\begin{equation}\label{eq:RelativeEntropy}
  D_\mathrm{KL}(t)
  \equiv \biggl\langle \ln \frac{\rho(\bm{x}_{\mspace{-1mu}t}, t)}{\pi \bl( \bm{x}_{\mspace{-1mu}t}, \lambda(t) \br)} \biggr\rangle
  = \langle -V(\bm{x}_{\mspace{-1mu}t}, t) \rangle
\end{equation}
quantifies the nonadiabatic lag between the state of the system at each time \( t \mspace{5mu}\mathord{\in}\mspace{5mu} [0, \tau] \) and its counterpart in the quasistatic limit~\cite{Sivak2012NearEquilibrium,Esposito2011SecondLaw}.
The second law of thermodynamics provides a decomposition of \( \Delta S \) through an entropy balance that clarifies \cref{eq:FreeEnergy}~\cite{Seifert2005EntropyProduction,Muratore-Ginanneschi2017EngineeredEquilibration}.
Since the system evolves at a fixed temperature, entropy changes in the ideal, implicit bath are reversible; we find (\nameref{sec:SI}, Section~II)
\begin{equation}\label{eq:Entropy}
  \Delta S
  = \biggl\langle \int_0^{\mspace{-1mu}\tau}\negthickspace \mathrm{d}t \thinspace D| \bm\nabla V(\bm{x}_{\mspace{-1mu}t}, t) |^2 \biggr\rangle - \beta \langle \mathcal{Q}_\tau \rangle
\end{equation}
where the first term is the entropy produced along the transformation, which is nonnegative, and the second is the dissipated heat introduced in \cref{eq:Energy}.
Combining \cref{eq:Energy,eq:FreeEnergy,eq:RelativeEntropy,eq:Entropy}, we finally obtain
\begin{equation}\label{eq:ForwardFreeEnergy}
  \Delta F = \langle \mathcal{W}_{\mspace{-2mu}\tau} \rangle - \beta^{-1} \biggl\langle \int_0^{\mspace{-1mu}\tau}\negthickspace \mathrm{d}t \thinspace D | \bm\nabla V(\bm{x}_{\mspace{-1mu}t}, t) |^2 - V(\bm{x}_{\mspace{-1mu}\tau}, \tau) \biggr\rangle
\end{equation}
an equality relating \( \Delta F \) to the dissipation along a nonadiabatic alchemical transformation from state \( \mathsf{A} \) toward state \( \mathsf{B} \).

\Cref{eq:ForwardFreeEnergy} furnishes a refinement of Jensen's bound on Jarzynski's equality in \cref{eq:Jarzynski},
\begin{equation}\label{eq:WorkBound}
  \Delta F \le \langle \mathcal{W}_{\mspace{-2mu}\tau} \rangle
\end{equation}
in that it explicitly states the dissipative terms that close the gap from Jensen's bound, rendering the inequality an equality.
These terms depend on the nonadiabatic potential \( V(\bm{x}, t) \) in \cref{eq:NonadiabaticPotential}, whose gradient can be regressed from the forward process by minimizing the loss function (\nameref{sec:SI}, Section~III)
\begin{equation}\label{eq:Loss}
  \ell (\bm{\theta})
  \equiv \biggl\langle \int_0^{\mspace{-1mu}\tau}\negthickspace \mathrm{d}t \thinspace D |\bm\nabla V_{\!\bm{\theta}}(\bm{x}_{\mspace{-1mu}t}, t)|^2 - 2 \negthinspace\int_0^{\mspace{-1mu}\tau}\negthickspace \mathrm{d}\bm{x}_{\mspace{-1mu}t} \mspace{2mu}\mathord{\circ}\mspace{2mu} \bm\nabla V_{\!\bm{\theta}}(\bm{x}_{\mspace{-1mu}t}, t) \biggr\rangle
\end{equation}
with respect to the parameters \( \bm{\theta} \) of the differentiable ansatz \( V_{\!\bm{\theta}} \).
If \( \bm{\theta}^\star \thinspace\mathord{\equiv}\thinspace \argmin \ell (\bm{\theta}) \) are optimal parameters, then the optimal nonadiabatic potential \( V_{\!\bm{\theta}^\star} \) satisfies
\begin{equation}\label{eq:OptimalNonadiabaticPotential}
  V(\bm{x}, t) = V_{\!\bm{\theta}^\star}(\bm{x}, t) - \ln \langle \exp[V_{\!\bm{\theta}^\star}(\bm{x}_{\mspace{-1mu}t}, t)] \rangle
\end{equation}
where the cumulant shift ensures correct normalization.
For the initial condition in \cref{eq:FokkerPlanck}, the optimal nonadiabatic potential must satisfy the boundary condition \( V_{\!\bm{\theta}^\star}(\bm{x}, t \smallequals 0) \thinspace\mathord{\equiv}\thinspace 0 \), which may be built into the ansatz or enforced through regularization during training.
While numerical estimation of the exponential average in Eq.~15 can be computationally challenging, it acts as a time-dependent baseline that can be incorporated into, and trained alongside, the ansatz for \( V(\boldsymbol{x}, t) \).

Taken together, \cref{eq:ForwardFreeEnergy,eq:Loss,eq:OptimalNonadiabaticPotential} suggest a scheme to estimate \( \Delta F \) by training an FMM to quantify the dissipation along a nonadiabatic alchemical transformation.
Given an ensemble of realizations of the forward process, a nonadiabatic potential ansatz \( V_{\!\bm{\theta}}(\bm{x}, t) \) can be trained by minimizing \cref{eq:Loss}, shifted according to \cref{eq:OptimalNonadiabaticPotential}, and plugged into \cref{eq:ForwardFreeEnergy} to obtain a free-energy estimate.
If the learned nonadiabatic potential is optimal, then this procedure yields a statistically unbiased estimate of \( \Delta F \) at no greater simulation cost than evaluating the work bound in \cref{eq:WorkBound}.

\section*{Flow matching models enable \\ variational free-energy estimation}

Since the exact nonadiabatic potential in \cref{eq:NonadiabaticPotential} is seldom known, unbiased free-energy estimation with \cref{eq:ForwardFreeEnergy} is all but intractable in realistic applications.
Here, we argue that the forward and backward processes that comprise an alchemical FMM are associated with variational upper and lower bounds on \( \Delta F \).
These bounds control the systematic bias incurred by a suboptimal nonadiabatic potential and, by virtue of their variational character, are simultaneously saturated by the equality in \cref{eq:ForwardFreeEnergy} for an optimal potential.

Probabilistic inference in FMMs is achieved through a map from noise to data that is enabled by a learned flow field.
In alchemical FMMs, this map is effected by the backward process
\begin{equation}\label{eq:BackwardProcess}
\begin{aligned}
  \mathrm{d}\bm{x}_{\mspace{-1mu}t} =
  &\medspace\mathord{-}D[ \beta \bm\nabla U \bl( \bm{x}_{\mspace{-1mu}t}, \lambda(\tau\mspace{-1mu}\smallminus\mspace{-1mu}t) \br) + 2 \bm\nabla V_{\!\bm{\theta}}(\bm{x}_{\mspace{-1mu}t}, \tau\mspace{-1mu}\smallminus\mspace{-1mu}t) ] \thinspace \mathrm{d}t \\[0.2em]
  &\mathbin{+}\sqrt{2D} \thinspace \mathrm{d}\bm{w}_{\mspace{-1mu}t}
\end{aligned}
\end{equation}
where the role of the flow field falls on the learned nonadiabatic force \( -\bm\nabla V_{\!\bm{\theta}} \).
A physical system evolving under the backward process has an initial configuration \( \bm{x}_{\mspace{-1mu}0} \) drawn from the noise distribution with density \( \rho(\bm{x}, \tau) \) and is driven by the time-reversed switch \( \lambda(\tau\mspace{-1mu}\smallminus\mspace{-1mu}t) \) toward the equilibrium distribution at state \( \mathsf{A} \).
If the learned nonadiabatic force is optimal, then the backward process maps the system directly onto state \( \mathsf{A} \), enabling the generation of equilibrium samples through the time-reversed nonadiabatic alchemical transformation.

A connection between the backward process and the free-energy difference \( \Delta F \) can be gleaned from the flexible trainability of FMMs through either flow matching or minimization of the relative entropy between forward and backward processes~\cite{Song2021MaximumLikelihood}.
Specifically, the nonadiabatic potential \( V_{\!\bm\theta} \) that enters the backward process can be trained by minimizing the loss in \cref{eq:Loss} or by minimizing the \emph{backward} Kullback--Leibler (KL) divergence
\begin{equation}\label{eq:PathKullbackLeiblerBackward}
  \biggl\langle
    \negthinspace\mathord{-} \ln \frac
    {\mathrm{d}\mathbb{P}_{\mspace{-5mu}\tau}[\{\bm{x}_{\mspace{-1mu}\tau-t}\}_{t=\mspace{1mu}0}^\tau]}
    {\mathrm{d}\mathbb{P}_{\mspace{-2mu}\bm{\theta},\tau}[\{\bm{x}_{\mspace{-1mu}t}\}_{t=\mspace{1mu}0}^\tau]}
  \biggr\rangle_{\negthickspace\bm{\theta}}
  \ge\thinspace 0
\end{equation}
between forward and backward path ensembles, where the symbol
\( \mathrm{d}\mathbb{P}_{\mspace{-5mu}\tau}[\{\bm{x}_{\mspace{-1mu}\tau-t}\}_{t=\mspace{1mu}0}^\tau] / \mathrm{d}\mathbb{P}_{\mspace{-2mu}\bm{\theta},\tau}[\{\bm{x}_{\mspace{-1mu}t}\}_{t=\mspace{1mu}0}^\tau] \)
denotes the likelihood of a backward trajectory \( \{\bm{x}_{\mspace{-1mu}t}\}_{t=\mspace{1mu}0}^\tau \) relative to that of its time-reversal \( \{\bm{x}_{\mspace{-1mu}\tau-t}\}_{t=\mspace{1mu}0}^\tau \), and the angled brackets \( \langle \,\cdot\, \rangle_{\bm\theta} \) denote an average over realizations.
The relative likelihood takes the form (\nameref{sec:SI}, Section~IV)
\begin{equation}\label{eq:PathRelativeDensityBackward}
  \frac
  {\mathrm{d}\mathbb{P}_{\mspace{-5mu}\tau}[\{\bm{x}_{\mspace{-1mu}\tau-t}\}_{t=\mspace{1mu}0}^\tau]}
  {\mathrm{d}\mathbb{P}_{\mspace{-2mu}\bm{\theta},\tau}[\{\bm{x}_{\mspace{-1mu}t}\}_{t=\mspace{1mu}0}^\tau]}
  = \exp\bigl[
    V(\bm{x}_{\mspace{-1mu}0}, \tau)
    - \beta \bigl( \overline{\mathcal{W}}_{\mspace{-2mu}\tau} + \Delta F \bigr)
    - \overline{\mathcal{S}}_{\bm{\theta},\tau}
  \bigr]
\end{equation}
where the reverse-time work and surprisal are given by
\begin{subequations}\label{eq:BackwardPathFunctionals}
\begin{align}
  \overline{\mathcal{W}}_{\mspace{-2mu}\tau}
  &\equiv \int_0^{\mspace{-1mu}\tau} \negthickspace \mathrm{d}t \thinspace \partial_t U \bl( \bm{x}_{\mspace{-1mu}t}, \lambda(\tau\mspace{-1mu}\smallminus\mspace{-1mu}t) \br)
  \label{eq:BackwardWork}
  \\ \overline{\mathcal{S}}_{\bm{\theta},\tau}
  &\equiv \int_0^{\mspace{-1mu}\tau}\negthickspace \mathrm{d}t \thinspace D |\bm\nabla V_{\!\bm{\theta}}(\bm{x}_{\mspace{-1mu}t}, \tau\mspace{-1mu}\smallminus\mspace{-1mu}t)|^2
  \label{eq:BackwardSurprisal}
  \\[-0.5em] &{\hskip 2.5em \relax}\mathord{-}\medspace \sqrt{2D} \thinspace \mathrm{d}\bm{w}_{\mspace{-1mu}t} \thinspace\mathord{\cdot}\thinspace \bm\nabla V_{\!\bm{\theta}}(\bm{x}_{\mspace{-1mu}t}, \tau\mspace{-1mu}\smallminus\mspace{-1mu}t)
  \nonumber
\end{align}
\end{subequations}
respectively.
Due to the nonnegativity of the KL divergence, one infers from \cref{eq:PathKullbackLeiblerBackward,eq:PathRelativeDensityBackward} the relation (\nameref{sec:SI}, Section~V~A)
\begin{equation}\label{eq:BackwardFreeEnergyBound}
\begin{aligned}
  \Delta F
  \ge &\medspace
  \langle \mathord{-}\overline{\mathcal{W}}_{\mspace{-2mu}\tau} \rangle_{\bm\theta}
  - \beta^{-1} \biggl\langle \int_0^{\mspace{-1mu}\tau}\negthickspace \mathrm{d}t \thinspace D| \bm\nabla V_{\!\bm{\theta}}(\bm{x}_{\mspace{-1mu}t}, \tau\mspace{-1mu}\smallminus\mspace{-1mu}t) |^2 \biggr\rangle_{\negthickspace\bm{\theta}}
  \\[0.25em] &\mathbin{+}
  \beta^{-1} \bigl[ \langle V_{\!\bm{\theta}}(\bm{x}_{\mspace{-1mu}0}, \tau) \rangle_{\bm\theta}
  - \ln \langle \exp[ V_{\!\bm{\theta}}(\bm{x}_{\mspace{-1mu}0}, \tau) ] \rangle_{\bm\theta} \bigr]
\end{aligned}
\end{equation}
which provides a \emph{backward variational bound} on \( \Delta F \) in terms of the work done and the dissipation incurred along the backward process.
A similar argument starting from the \emph{forward} KL divergence between forward and backward processes yields the separate relation (\nameref{sec:SI}, Section~V~B)
\begin{equation}\label{eq:ForwardFreeEnergyBound}
\begin{aligned}
  \Delta F
  \le
  &\medspace
  \langle \mathcal{W}_{\mspace{-2mu}\tau} \rangle -
  \beta^{-1} \biggl\langle \int_0^{\mspace{-1mu}\tau}\negthickspace \mathrm{d}t \thinspace D| \bm\nabla V_{\!\bm{\theta}}(\bm{x}_{\mspace{-1mu}t}, t) |^2 \biggr\rangle
  \\ &\mathbin{+} \beta^{-1} \bigl[ \langle V_{\!\bm{\theta}}(\bm{x}_{\mspace{-1mu}\tau}, \tau) \rangle - \ln \langle \exp[V_{\!\bm{\theta}}(\bm{x}_{\mspace{-1mu}\tau}, \tau)] \rangle \bigr]
\end{aligned}
\end{equation}
and thus recasts \cref{eq:ForwardFreeEnergy} as the saturated case of a \emph{forward variational bound} on the free-energy difference.

\begin{figure}[ht]
\centering
\includegraphics[width=2.8in]%
  {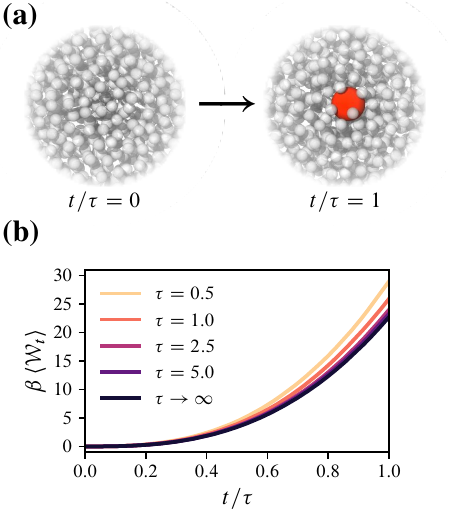}
\caption{%
  Alchemical solvation in a WCA liquid.
  \textbf{(a)}
  Configuration snapshots along the alchemical path of a growing solute (red) within a WCA solvent (gray).
  \textbf{(b)}
  Estimates of the cumulative work done on the system to grow the solute at various rates \( \tau^{-1} \), compared to adiabatic growth at the quasistatic limit (black).
}\label{fig:wca_solvation_bare_work}
\end{figure}

The variational lower and upper bounds on \( \Delta F \) in \cref{eq:BackwardFreeEnergyBound,eq:ForwardFreeEnergyBound} are both saturated by the optimal nonadiabatic potential \( V_{\!\bm{\theta}^\star}(\bm{x}, t) \) that satisfies \cref{eq:OptimalNonadiabaticPotential} (\nameref{sec:SI}, Section~IV).
Away from saturation, these inequalities provide a confidence interval for the free-energy difference that can be systematically narrowed through refinement of the nonadiabatic potential ansatz.
It is worth noting that a free-energy estimate obtained by evaluating the backward variational bound in \cref{eq:BackwardFreeEnergyBound} can itself be systematically improved through the application of reinforcement learning schemes to saturate the inequality~\cite{Hartmann2017VariationalCharacterization,Das2021ReinforcementLearning}.

\section*{Efficiency gains from nonadiabatic flow matching for free-energy estimation}

We provide a numerical demonstration of alchemical FMMs for free-energy estimation by evaluating the backward variational bound in \cref{eq:BackwardFreeEnergyBound} on atomistic test problems.
We use a simple linear switch \( \lambda(t) \medspace\mathord{\equiv}\medspace (1 - t / \tau) \thinspace  \lambda_\mathsf{A} + (t / \tau) \thinspace \lambda_\mathsf{B} \), where \( \lambda_\mathsf{A} \thinspace\mathord{\equiv}\thinspace 0 \) and \( \lambda_\mathsf{B} \thinspace\mathord{\equiv}\thinspace 1 \) respectively index the equilibrium states \( \mathsf{A} \) and \( \mathsf{B} \) at the endpoints of the alchemical transformation.
This switch recovers the quasistatic limit as \( \tau \thinspace\mathord{\to}\thinspace \infty \), allowing us to compare the simulation cost to evaluate the backward variational bound in \cref{eq:BackwardFreeEnergyBound} with that of the TI estimator in \cref{eq:TI}.
We set the diffusion coefficient to unity and use the Euler--Maruyama integrator~\cite{Kloeden1992Book} with integration stepsizes up to \( 5 \times 10^{-5} \) time units to estimate forward and backward trajectories by solving \cref{eq:ForwardProcess} and \cref{eq:BackwardProcess}, respectively.

\subsection*{WCA monomer solvation}

As a canonical benchmark problem, we consider the free energy to solvate a point particle in a three-dimensional Weeks--Chandler--Andersen (WCA) solvent.
In our simulations, the solvent is comprised by \( N \thinspace\mathord{=}\thinspace 512 \) point particles with positions \( \{\bm{x}_{\mspace{-1mu}i}\}_{i=1}^{N} \), their interactions governed by the pairwise energy function
\begin{equation}\label{eq:WCA}
  u_\mathrm{WCA}(\bm{x}_{\mspace{-1mu}i}, \bm{x}_{\mspace{-1mu}j} |\mspace{2mu} \sigma)
  \equiv 4\epsilon \thinspace ( r_{ij}^{-12} - r_{ij}^{-6} + 1/4 ) \thinspace \bm{1}\{ r_{ij} \!\le 2^{1/6} \}
\end{equation}
with \( \sigma \) the solvent size, \( \epsilon \) the interaction energy scale, \( r_{ij} \thinspace\mathord{\equiv}\thinspace |\bm{x}_{\mspace{-1mu}j} - \bm{x}_{\mspace{-1mu}i}| / \sigma \) the dimensionless distance between particles \( i \) and \( j \), and \( \bm{1} \{ r_{ij} \thinspace\mathord{\le}\thinspace \,\cdot\, \} \) a distance cutoff indicator.
The solvent occupies a periodic cubic box with a side length of \( 8.110 \, \sigma \) and is held at a temperature that corresponds to \( \beta^{-1} \thinspace\mathord{=}\thinspace 1.214 \, \epsilon \).
These parameters define the equilibrium distribution of the system at state \( \mathsf{A} \), whose density is proportional to the Boltzmann factor \( \smash{\exp \bigl[ -\beta \sum\nolimits_{i < j} u_\mathrm{WCA}(\bm{x}_{\mspace{-1mu}i}, \bm{x}_{\mspace{-1mu}j} |\mspace{2mu} \sigma) \bigr]} \), where the sum runs over solvent atom pairs.

The alchemical path to the solvated state starts with the insertion of a small solute into a random pocket of the equilibrium solvent.
The free-energetic cost of this operation, which we neglect, is well captured by a Poisson point process approximation of Widom's scheme that takes advantage of the solute's small size at insertion~\cite{Widom1963SomeTopics,Crooks1997GaussianStatistics}.
Once inserted, the solute grows from its initial size, \( \sigma_\mathsf{A} \thinspace\mathord{\equiv}\thinspace 0.02 \, \sigma \), to its final size, \( \sigma_\mathsf{B} \thinspace\mathord{\equiv}\thinspace 2 \, \sigma \), while interacting with the surrounding solvent.
For this $\sigma_\mathsf{A}$, the expected free energy from Widom insertion is much less than the thermal energy.
Solute-solvent interactions are also governed by the energy function in \cref{eq:WCA}, but involve a switch-dependent interaction lengthscale
\begin{equation}\label{eq:GrowthSchedule}
  \sigma_\mathrm{S}(t)
  \equiv \sqrt{[\smash{ \sigma_\mathsf{A} \thinspace \bl( 1 - \lambda(t) \br) + \sigma_\mathsf{B} \thinspace \lambda(t) }] \thinspace \sigma}
\end{equation}
that dictates the solute size.
With the solute position given by \( \bm{x}_\mathrm{S} \), the switching potential takes the form
\begin{equation}\label{eq:GrowthPotential}
\begin{aligned}
  U \bl( \bm{x}, \lambda(t) \br) \equiv
  &\thickspace\sum\nolimits_{i < j}
  u_\mathrm{WCA}(\bm{x}_{\mspace{-1mu}i}, \bm{x}_{\mspace{-1mu}j} |\mspace{2mu} \sigma)
  \\[0.5em]
  &\mathbin{+}\sum\nolimits_{i}
  u_\mathrm{WCA}\bl(\bm{x}_{\mspace{-1mu}i}, \bm{x}_\mathrm{S} |\mspace{2mu} \sigma_\mathrm{S}(t)\br)
\end{aligned}
\end{equation}
where \( \bm{x} \medspace\mathord{\equiv}\medspace \vector \bl( \{ \bm{x}_\mathrm{S} \}, \{ \bm{x}_{\mspace{-1mu}i} \}_{i=1}^N \br) \) is the solvated configuration.
The alchemical path to the solvated state is illustrated in \cref{fig:wca_solvation_bare_work}a, which contains simulation snapshots taken along a solute growth trajectory obtained from the forward process in \cref{eq:ForwardProcess} with the potential in \cref{eq:GrowthPotential}.

\Cref{fig:wca_solvation_bare_work}b shows Monte Carlo estimates of the mean accumulated work \( \langle \mathcal{W}_t \rangle \) done on the system to enact solute growth up to each time \( t \in [0, \tau] \) during the nonadiabatic alchemical transformation.
The work is computed at growth rates \( \tau^{-1} \) spanning an order of magnitude, and separately for a quasistatic growth rate via trapezoidal quadrature of \cref{eq:TI} on a \( 21 \)-point grid along the switching schedule with a uniform grid spacing of \( 0.05 \).
The cumulative work estimates, obtained by averaging \( 100 \) Euler--Maruyama solutions of \cref{eq:ForwardProcess} at each growth rate, exhibit negligible variance and accrue smoothly along the thermodynamic path, suggesting that the excess work, \( \langle \mathcal{W}_t \rangle \thinspace\mathord{-}\medspace [ F\bl(\lambda(t)\br) - F\bl(\lambda(0)\br) ] \), arises from localized energy fluctuations throughout the growth process.
The leading-order role of this excess work is to rectify solvation-shell fluctuations that contend with solute growth.

\begin{figure}[t]
\centering
\includegraphics[width=2.8in]%
  {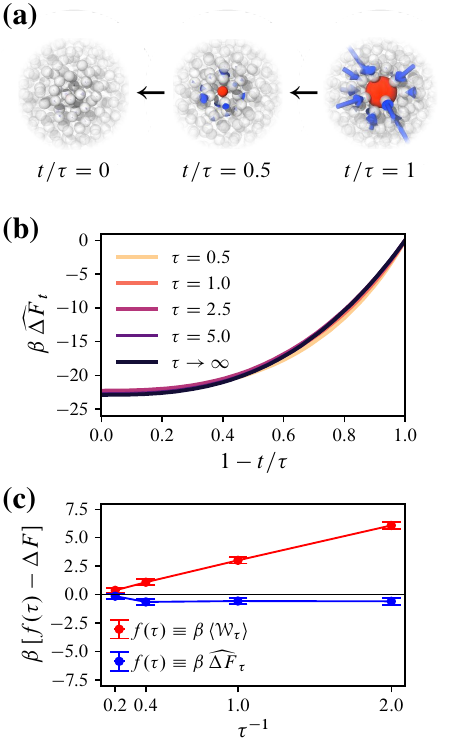}
\caption{%
  WCA solvation free energies from nonadiabatic force matching.
  \textbf{(a)}
  Snapshots from a realization of the backward process, where the solute growth is reversed by the nonadiabatic switching potential.
  \textbf{(b)}
  The cumulative free-energy differences estimated from the denoised solute growth (colored) closely match the quasistatic limit (black).
  \textbf{(c)}
  Accuracy of solvation free-energy estimates as a function of solute growth rate \( \tau^{-1} \).
}\label{fig:wca_solvation_variational_dissipation}
\end{figure}

In general, physical insight into the role of excess work informs the design of efficiently trainable nonadiabatic potential ansätze for accurate free-energy estimation.
We use an ansatz that captures energetic fluctuations due to solute-solvent interactions along the solute growth direction while ignoring higher-order contributions due to solvent-solvent interactions.
The ansatz is comprised of a Behler-style neural forcefield~\cite{Behler2007BPNN} whose inputs are \( N \) solute-solvent pair distances \( \{ \phi(|\bm{x}_\mathrm{S} \thinspace\mathord{-}\medspace \bm{x}_{\mspace{-1mu}i}| / \sigma) \}_{i=1}^{N} \), where
\begin{equation}
  \phi(r) \equiv
  \frac{1}{2} [ 1 + \cos ( \pi r^2 \negthinspace / 4 ) ] \exp( -r^2 \negthinspace / 4 ) \thinspace \bm{1}\{ r \le 2 \}
\end{equation}
is a symmetry function that encodes short-ranged, rotationally invariant, repulsive interactions between the solute and the solvent~\cite{Behler2011AtomCentered}.
These inputs are passed to a shallow, fully-connected neural network (NN) with \( N \) neurons in the input layer, \( 4 N \) neurons in a hidden layer, and a scalar output layer that yields the nonadiabatic potential \( V_{\!\bm{\theta}}(\bm{x}, t) \).
To ensure that the learned potential satisfies the boundary condition \( V_{\!\bm{\theta}}(\bm{x}, t \smallequals 0) \equiv 0 \), the NN inputs are multiplied by the instantaneous switching value \( \lambda(t) \), the linear bias of each layer is fixed at zero, and exponential linear units~\cite{Clevert2016ELU} are used as activation functions in all layers.
The layer weights \( \bm{\theta} \) are optimized by minimizing the loss in \cref{eq:Loss} with the Adam optimizer~\cite{Kingma2015Adam}, and the optimal nonadiabatic potential \( V_{\!\bm{\theta}^\star}(\bm{x}, t) \) is used to produce \( 100 \) statistically independent Euler--Maruyama solutions of \cref{eq:BackwardProcess} at each growth rate, each initialized at the final configuration along a corresponding solution of \cref{eq:ForwardProcess}.

Snapshots taken along a trajectory of the learned backward process are shown in \cref{fig:wca_solvation_variational_dissipation}a, where the blue arrows show how the optimal nonadiabatic force \( -\nabla V_{\!\bm{\theta}^\star}(\bm{x}, t) \) drives solvent particles to fill the vacancy left by the solute as it shrinks in reverse time.
By construction, this force is localized around the solute's closest solvation shell throughout the switching process. Although this may result in a reasonable approximation of the fluctuating velocity field near the quasistatic limit, it rapidly degrades as the field delocalizes at solute growth rates faster than those considered here.
A total of \( 100 \) independent backward trajectories are used to evaluate an estimator of \cref{eq:BackwardFreeEnergyBound},
\begin{equation}\label{eq:BackwardFreeEnergyBoundEstimator}
  \widehat{\Delta F}_t
  \equiv
  \langle -\overline{\mathcal{W}}_t \rangle_{\bm\theta}
  - \beta^{-1}\langle \overline{\mathcal{S}}_{\bm\theta,t} \rangle_{\bm\theta}
  + \beta^{-1}\langle V_{\!\bm\theta}(\bm{x}_{\mspace{-1mu}t}, \tau\mspace{-1mu}\smallminus\mspace{-1mu}t) \rangle_{\bm\theta}
\end{equation}
for times \( t \) within the duration of the nonadiabatic alchemical transformation.
This estimator excludes the cumulant shift in \cref{eq:BackwardFreeEnergyBound}, which we found to be negligible in our calculations.
The results, shown in \cref{fig:wca_solvation_variational_dissipation}b, illustrate that the variational free-energy bound is closely saturated across the range of transition durations.
Comparing the accuracy of the variational bound in \cref{eq:BackwardFreeEnergyBound} with the work bound in \cref{eq:WorkBound}, as done in \cref{fig:wca_solvation_variational_dissipation}c, shows that the trajectory duration required to saturate the variational free-energy estimate is between \( 1 \)-\( 2 \) orders of magnitude shorter than that required to approach the quasistatic limit, which in turn suggests that the backward variational bound attains a commensurate reduction in simulation cost relative to TI.

\subsection*{LJ solid formation}

To illustrate the utility of flow matching for free-energy estimation in solids, we compute the free energy to form a Lennard--Jones (LJ) solid from a harmonic face-centered cubic (FCC) crystal.
We simulate a periodic super-lattice with \( 5 \thinspace\mathord{\times}\thinspace 5 \thinspace\mathord{\times}\thinspace 5 \) FCC unit cells or, equivalently, \( N \thinspace\mathord{=}\thinspace 500 \) lattice sites with positions \( \bm{x}_\mathrm{O} \thickspace\mathord{\equiv}\thinspace \vector ( \{ \bm{x}_{\mathrm{O},i} \}_{i=1}^{N} ) \), of which one is kept frozen to fix the center of mass of the lattice~\cite{Vega2007EinsteinMolecule}.
The unit cell side-length is \( 1.462 \, \sigma \), where \( \sigma \) sets the interaction lengthscale, and the lattice is held at a temperature that corresponds to \( \beta^{-1} \thinspace\mathord{=}\medspace \epsilon \), where \( \epsilon \) sets the interaction energy scale.
The harmonic FCC lattice is alchemically evolved toward a LJ solid via the Frenkel--Ladd potential~\cite{Frenkel1984EinsteinCrystal}
\begin{equation}\label{eq:FormationPotential}
  U \bl( \bm{x}, \lambda(t) \br)
  \equiv
  \frac{\kappa}{2} \bl( 1 - \lambda(t) \br) \thinspace |\mspace{1mu}\bm{x} - \bm{x}_\mathrm{O}|^2
  + \lambda(t) \thinspace U_\mathrm{LJ}(\bm{x})
\end{equation}
where \( \bm{x} \medspace\mathord{\equiv}\thinspace \vector ( \{ \bm{x}_{\mspace{-1mu}i} \}_{i=1}^{N} ) \) is the fluctuating configuration of the lattice, \( U_\mathrm{LJ}(\bm{x}) \thickspace\mathord{\equiv}\thickspace \smash{\sum\nolimits_{i < j}} \thinspace u_\mathrm{LJ}(\bm{x}_{\mspace{-1mu}i}, \bm{x}_{\mspace{-1mu}j}) \) the LJ potential with the pairwise energy function
\begin{equation}\label{eq:LJ}
  u_\mathrm{LJ}(\bm{x}_{\mspace{-1mu}i}, \bm{x}_{\mspace{-1mu}j})
  \equiv 4\epsilon \thinspace [ (|\bm{x}_{\mspace{-1mu}j} - \bm{x}_{\mspace{-1mu}i}|/\sigma)^{-12} - (|\bm{x}_{\mspace{-1mu}j} - \bm{x}_{\mspace{-1mu}i}|/\sigma)^{-6} ]
\end{equation}
truncated and shifted to zero at pairwise distances \( |\bm{x}_{\mspace{-1mu}j} - \bm{x}_{\mspace{-1mu}i} | \ge 2.7 \, \sigma \)~\cite{Frenkel2002Book},
and \( \kappa \thinspace\mathord{\equiv}\thinspace 100 \, \epsilon / \sigma^2 \) the stiffness constant of the harmonic crystal.

\begin{figure}[t]
\centering
\includegraphics[width=2.8in]%
  {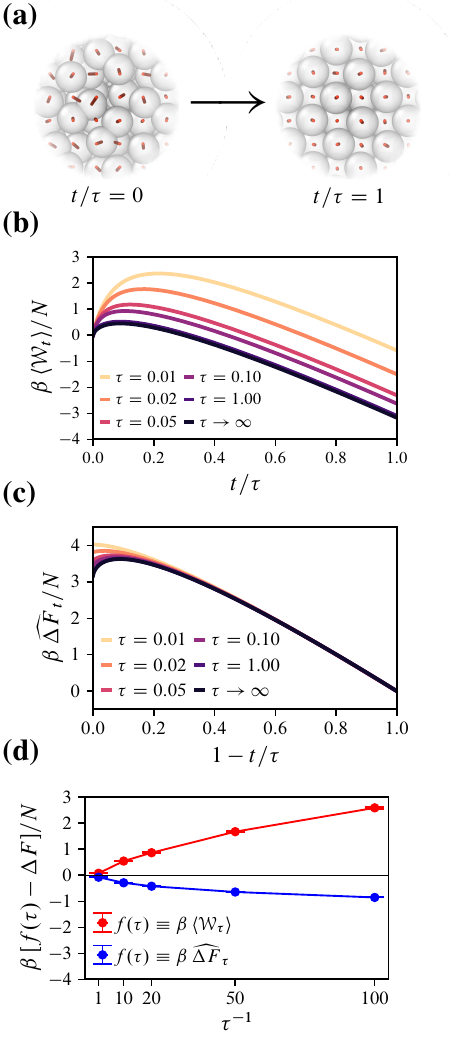}
\caption{%
  Estimates of the free energy to form an LJ solid.
  \textbf{(a)}
  Configuration snapshots along the Frenkel--Ladd alchemical path, wherein a harmonic FCC lattice is driven toward an LJ solid.
  \textbf{(b)}
  The cumulative work per site at lattice relaxation rates \( \tau^{-1} \) spanning two orders of magnitude.
  \textbf{(c)}
  Backward estimates of the formation free-energy difference per site are accurate across a range of relaxation rates \( \tau^{-1} \).
  \textbf{(d)}
  The bias in the formation free-energy difference per site at alchemical switching rates \( \tau^{-1} \).
}\label{fig:fcc_formation}
\end{figure}

\Cref{fig:fcc_formation}a gives trajectory snapshots for the alchemical transformation from a harmonic crystal to the LJ lattice, with the red lines indicating the displacement of each particle from its FCC lattice site.
Monte Carlo estimates of the work done on the system \emph{per lattice site}, \( \langle \mathcal{W}_t \rangle / N \), up to each time \( t \in [0, \tau] \) are shown in \cref{fig:fcc_formation}b for switching rates \( \tau^{-1} \) spanning two orders of magnitude.
At each lattice relaxation rate, \( 100 \) statistically independent Euler--Maruyama solutions of the forward process in \cref{eq:ForwardProcess} are used to estimate the cumulative work done on the system, which stores excess energy in lattice modes that cannot be dispersed within the timescale of the alchemical transformation.
The cumulative work in the quasistatic limit as \( \tau^{-1} \thinspace\mathord{\to}\medspace 0 \), also shown in \cref{fig:fcc_formation}b, was approximated by averaging over trajectories of duration \( \tau = 10 \), resulting in a ground-truth estimate of \( \Delta F \) that is comparable in accuracy to TI with a fine quadrature grid.
For this system, the free-energy difference is well-approximated by the cumulative work [\cref{eq:WorkBound}] at remarkably short durations.
Because the fluctuating lattice configuration remains sharply concentrated around the mode of the adiabatic alchemical distribution, nonadiabatic fluctuations are rendered a higher-order effect.

This observation has important consequences for variational free-energy estimation as shown in \cref{fig:fcc_formation}c, where the estimator in \cref{eq:BackwardFreeEnergyBoundEstimator} is evaluated with the \emph{constant} nonadiabatic potential ansatz \( V_{\!\bm{\theta}}(\bm{x}, t) \equiv 0 \) using \( 100 \) Euler--Maruyama solutions of the backward process in \cref{eq:BackwardProcess}, initialized with the final nonequilibrium distribution sampled by the forward trajectories.
In alignment with our expectation that the exact nonadiabatic potential is roughly constant along the nonadiabatic alchemical transformation, the estimator in \cref{eq:BackwardFreeEnergyBoundEstimator} with a constant potential closely follows the variational free-energy bound in \cref{eq:BackwardFreeEnergyBound} across a wide range of transition durations.
In contrast, the work bound in \cref{eq:WorkBound} is approximately thrice as biased across the range of alchemical switching rates \( \tau^{-1} \), as seen in \cref{fig:fcc_formation}d.
Thus in this example, simply using the backward variational bound without training an FMM significantly reduces the error of nonadiabatic free-energy estimation.

\section*{Concluding remarks}

We present a stochastic-thermodynamic framework for free-energy estimation along a nonadiabatic alchemical transformation that hinges on training a flow-based generative model.
Within this framework, we derive variational estimators for the free-energy difference that, coupled with an efficient neural ansatz for the nonadiabatic force field, can reduce the simulation cost of a statistically unbiased free-energy estimate by orders of magnitude relative to methods that operate within the quasistatic limit, such as thermodynamic integration.
In our applications of the variational free-energy bound in \cref{eq:BackwardFreeEnergyBound}, we use computationally inexpensive ansätze for the nonadiabatic force to resolve its fluctuations along dominant dissipation pathways while ignoring higher-order interaction channels that contribute negligibly to the overall dissipation.
In doing so, we illustrate the benefits of using system-adapted neural ansätze to more efficiently evaluate alchemical transitions.
All numerical results presented in this work, along with generating code, are available in the GitHub repository~\href{https://github.com/jrosaraices/nonadiabatic-force-matching}{\texttt{jrosaraices/nonadiabatic-force-matching}}.

The current work adds to a rapidly growing collection of free-energy estimation schemes that leverage controlled diffusions.
Of note are Máté~\emph{et.~al.}'s work on \emph{Neural~TI}~\cite{Mate2024NeuralTI,Mate2025SolvationFE}, and He~\emph{et.~al.}'s work on free-energy difference estimation via adaptive transport~\cite{He2025FEAT,He2025RNE}.
In these works, a counterdiabatic potential is sought to advect a molecular system along a set of equilibrium states connecting \( \mathsf{A} \) and \( \mathsf{B} \), in the spirit of the \emph{escorted work} framework introduced by Vaikuntanathan and Jarzynski~\cite{Vaikuntanathan2008EscortedFE,Zhong2024TimeAsymmetric}.
In contrast, our approach to free-energy estimation evolves from Hummer's \emph{fast growth} framework~\cite{Hummer2001FastGrowth}, to which we add flow matching to mitigate biases associated with estimating the moment-generating function in Jarzynski's equality.
Leaving an explicit performance comparison for the future, we believe that the fast-growth approach exhibits a computational advantage over the escorted approach in settings where training-adapted data generation would be too expensive or done \emph{ex silico}, or where simulation-free training would bypass insurmountable cost bottlenecks~\cite{He2025SimulationFree}.

\newpage

\section*{Supporting Information}\setcurrentname{SI}\label{sec:SI}

Mathematical derivation of the free-energy difference along a nonadiabatic alchemical transformation [\cref{eq:FreeEnergy}]; development of the entropy balance along a diffusive process [\cref{eq:Entropy}]; justification of the loss function to train the nonadiabatic force [\cref{eq:Loss}]; derivation of the relative path density between backward and forward diffusive processes [\cref{eq:PathRelativeDensityBackward}]; derivation of the backward and forward bounds on the free-energy difference [\cref{eq:BackwardFreeEnergyBound,eq:ForwardFreeEnergyBound}]; and proof that the free-energy bounds are saturated by the optimal nonadiabatic potential

\section*{Acknowledgments}

This work was supported by the U.S.\ Department of Energy, Office of Science, Office of Advanced Scientific Computing Research, and Office of Basic Energy Sciences, via the Scientific Discovery through Advanced Computing (SciDAC) program under Award Number DE-SC0022364.
J.L.R.-R.\ acknowledges support from the National Science Foundation through the MPS-Ascend Postdoctoral Research Fellowship, Award No.\ \( 2213064 \).
The authors thank Yuanqi Du and Aditya N.\ Singh for stimulating discussions, and Bingqing Cheng for helpful comments on an early version of this article.

\bibliography{main}

\end{document}


\title{%
  Supporting Information: \emph{%
  Nonadiabatic force matching \\
  for alchemical free-energy estimation%
}}

\author{Jorge L.\ Rosa-Raíces}
\affiliation{%
  Department of Chemistry,
  University of California,
  Berkeley, California 94720, USA}

\author{David T.\ Limmer}
\email{dlimmer@berkeley.edu}
\affiliation{%
  Department of Chemistry,
  University of California,
  Berkeley, California 94720, USA}
\affiliation{%
  Materials Science Division,
  Lawrence Berkeley National Laboratory,
  Berkeley, California 94720, USA}
\affiliation{%
  Chemical Science Division,
  Lawrence Berkeley National Laboratory,
  Berkeley, California 94720, USA}
\affiliation{%
  Kavli Energy NanoScience Institute,
  Berkeley, California 94720, USA}

\maketitle

This document supplements the Main Text with mathematical derivations of results presented therein.
Reference to the following sections are placed at the relevant locations in the Main Text.

\section{Free-energy difference along a nonadiabatic alchemical transformation}
\label{sec:FreeEnergyDerivation}

Here, we provide a detailed derivation of \cref{eq:FreeEnergy} of the main text, which corresponds to the Helmholtz free-energy difference accrued along a nonadiabatic alchemical transformation.
Starting from the first law of thermodynamics as stated in \cref{eq:Energy} of the main text, we compute:
\begin{subequations}
\begin{align}
  \Delta U
  &\equiv \langle U \bl( \bm{x}_{\mspace{-1mu}\tau}, \lambda(\tau) \br) - U \bl( \bm{x}_{\mspace{-1mu}0}, \lambda(0) \br) \rangle
  \nonumber \\[5pt]
  &= F \bl( \lambda(\tau) \br) - F \bl( \lambda(0) \br)
    - \beta^{-1} \langle \ln \pi \bl( \bm{x}_{\mspace{-1mu}0}, \lambda(0) \br) - \ln \pi \bl( \bm{x}_{\mspace{-1mu}\tau}, \lambda(\tau) \br) \rangle
  \label{seq:FreeEnergy1} \\[5pt]
  &\equiv \Delta F
    - \beta^{-1} \langle \ln \pi( \bm{x}_{\mspace{-1mu}0}, \lambda_\mathsf{A} ) - \ln \pi( \bm{x}_{\mspace{-1mu}\tau}, \lambda_\mathsf{B} ) \rangle
  \label{seq:FreeEnergy2} \\
  &= \Delta F
    - \beta^{-1} \biggl[ \langle \ln \rho( \bm{x}_{\mspace{-1mu}0}, 0 ) - \ln \rho( \bm{x}_{\mspace{-1mu}\tau}, \tau ) \rangle
    +\mspace{3mu} \biggl\langle \ln\frac{\rho( \bm{x}_{\mspace{-1mu}\tau}, \tau)}{\pi \bl( \bm{x}_{\mspace{-1mu}\tau}, \lambda(\tau) \br)} \biggr\rangle \biggr]
  \nonumber \\
  &\equiv \Delta F
    - \beta^{-1} \bigl[ \Delta S + D_\mathrm{KL}(\tau) \bigr]
  \label{seq:FreeEnergy3}
\end{align}
\end{subequations}
where we rearranged the equilibrium density \( \pi(\bm{x}, \lambda) \) introduced in \cref{eq:EquilibriumDensity} of the main text
to obtain the second equality [\cref{seq:FreeEnergy1}]; the definition
\(
  \Delta F
  \equiv
  F \bl( \lambda(\tau) \br)
  -
  F \bl( \lambda(0) \br)
\)
along with the protocol boundary values
\(
  \lambda(t \smallequals 0)
  \equiv
  \lambda_\mathsf{A}
\)
and
\(
  \lambda(t \smallequals \tau)
  \equiv
  \lambda_\mathsf{B}
\)
to obtain the third equality [\cref{seq:FreeEnergy2}]; and the definitions of \( \Delta S \) and \( D_\mathrm{KL}(\tau) \) stated in the main text, namely
\begin{equation}\label{seq:EntropyDefinitions}
  \Delta S \equiv
  \biggl\langle \negthinspace\mathord{-}\ln \frac{\rho(\bm{x}_{\mspace{-1mu}\tau}, \tau)}{\rho(\bm{x}_{\mspace{-1mu}0}, 0)} \biggr\rangle
  \quad \text{and} \quad
  D_\mathrm{KL}(\tau) \equiv
  \biggl\langle \ln \frac{\rho(\bm{x}_{\mspace{-1mu}\tau}, \tau)}{\pi \bl(\bm{x}_{\mspace{-1mu}\tau}, \lambda(\tau) \br)} \biggr\rangle
\end{equation}
to finally obtain \cref{seq:FreeEnergy3}, which is the Helmholtz free-energy difference in \cref{eq:FreeEnergy} of the main text.

\section{Entropy balance along a diffusive process}
\label{sec:EntropyDerivation}

Here we derive \cref{eq:Entropy} of the main text, which clarifies the dynamics of entropy production along a diffusive process that is driven by an implicit bath at constant temperature.
To begin, we note that the entropy change \( \Delta S \) defined in \cref{seq:EntropyDefinitions} can be expanded into
\begin{equation}\label{seq:StratonovichChainRule}
  \Delta S
  \medspace=\medspace
  \biggl\langle \negthinspace\mathord{-}\int_0^{\mspace{-1mu}\tau} \negthinspace \mathrm{d} \ln \rho (\bm{x}_{\mspace{-1mu}t}, t) \biggr\rangle
  \medspace=\medspace
  \biggl\langle \negthinspace\mathord{-}\int_0^{\mspace{-1mu}\tau} \negthinspace \mathrm{d}t \thinspace \partial_t \ln \rho(\bm{x}, t) \biggr\rangle
  \medspace+\medspace
  \biggl\langle \negthinspace\mathord{-}\int_0^{\mspace{-1mu}\tau} \negthinspace \mathrm{d}\bm{x}_{\mspace{-1mu}t} \circ \nabla \ln \rho(\bm{x}_{\mspace{-1mu}t}, t) \biggr\rangle
\end{equation}
where we used the \emph{Stratonovich chain rule} (see, e.g., Proposition~3.4 in Ref.~\onlinecite{Pavliotis2014Book}) to expand an integral along the total differential \( \mathrm{d} \ln \rho(\bm{x}_{\mspace{-1mu}t}, t) \) into a deterministic and a (Stratonovich) stochastic integral.
The deterministic integral vanishes upon an application of Leibniz's integral rule,
\begin{align}
  \biggl\langle \int_0^{\mspace{-1mu}\tau} \negthinspace \mathrm{d}t \, \partial_t \ln \rho (\bm{x}_{\mspace{-1mu}t}, t) \biggr\rangle
  &=
  \int_0^{\mspace{-1mu}\tau} \negthinspace \mathrm{d}t \, \langle \partial_t \ln \rho (\bm{x}_{\mspace{-1mu}t}, t) \rangle
  =
  \int_0^{\mspace{-1mu}\tau} \negthinspace \mathrm{d}t \int \mathrm{d}\bm{x} \thinspace \rho(\bm{x}, t) \thinspace \partial_t \ln \rho(\bm{x}, t)
  \nonumber \\[5pt] &=
  \int_0^{\mspace{-1mu}\tau} \negthinspace \mathrm{d}t \int \mathrm{d}\bm{x} \thinspace \partial_t \rho(\bm{x}, t)
  =
  \int_0^{\mspace{-1mu}\tau} \negthinspace \mathrm{d}t \thinspace \biggl[ \frac{\hfill\mathrm{d}}{\hfill\mathrm{d}t} \int \mathrm{d}\bm{x} \thinspace \rho (\bm{x}, t) \biggr]
  \nonumber \\[5pt] &=
  \int_0^{\mspace{-1mu}\tau} \negthinspace \mathrm{d}t \thinspace \biggl[ \frac{\hfill\mathrm{d}}{\hfill\mathrm{d}t} \thinspace \text{const.} \biggr]
  =
  0
  \label{seq:StratonovichChainRuleDeterministic}
\end{align}
where the third equality follows from our assumption that the probability density \( \rho(\bm{x}, t) \) is positive for all \( \bm{x} \) at each time \( t \in [0, \tau] \), and the penultimate equality follows from the fact that probability is conserved along the diffusive process.
The stochastic integral in \cref{seq:StratonovichChainRule} is rendered deterministic by the equality
\begin{equation}\label{seq:StratonovichIdentity}
  \biggl\langle \int_0^{\mspace{-1mu}\tau} \negthinspace \mathrm{d}\bm{x}_{\mspace{-1mu}t} \circ \bm{f}(\bm{x}_{\mspace{-1mu}t}, t) \biggr\rangle
  = \biggl\langle \int_0^{\mspace{-1mu}\tau} \negthickspace \mathrm{d}t \thinspace D \bm\nabla V(\bm{x}_{\mspace{-1mu}t}, t) \cdot \bm{f}(\bm{x}_{\mspace{-1mu}t}, t) \biggr\rangle
\end{equation}
with \( V(\bm{x}, t)  \) the \emph{nonadiabatic potential} defined in \cref{eq:NonadiabaticPotential} of the main text.
This equality is valid for any sufficiently smooth vector-valued function \( \bm{f} \), and is derived in \cref{sec:LossDerivation} of this document.
Using \cref{seq:StratonovichIdentity}, we rewrite the second term on the right-hand side of \cref{seq:StratonovichChainRule} as
\begin{subequations}\label{seq:StratonovichChainRuleStochastic}
\begin{align}
  \biggl\langle
    \negthinspace\mathord{-}\int_0^{\mspace{-1mu}\tau} \negthinspace \mathrm{d}\bm{x}_{\mspace{-1mu}t} \circ \bm\nabla \ln \rho(\bm{x}_{\mspace{-1mu}t}, t)
  \biggr\rangle
  &= \biggl\langle
    \negthinspace\mathord{-}\int_0^{\mspace{-1mu}\tau} \negthickspace \mathrm{d}t \thinspace D \bm\nabla V(\bm{x}_{\mspace{-1mu}t}, t) \cdot \bm\nabla \ln \rho(\bm{x}_{\mspace{-1mu}t}, t)
  \biggr\rangle
  \label{seq:StratonovichChainRuleStochastic1} \\
  &= \biggl\langle
    \int_0^{\mspace{-1mu}\tau} \negthickspace \mathrm{d}t \thinspace D \bm\nabla V(\bm{x}_{\mspace{-1mu}t}, t) \cdot [ \bm\nabla V(\bm{x}_{\mspace{-1mu}t}, t) - \bm\nabla \ln \pi \bl( \bm{x}_{\mspace{-1mu}t}, \lambda(t) \br)]
  \biggr\rangle
  \nonumber \\
  &= \biggl\langle
    \int_0^{\mspace{-1mu}\tau} \negthickspace \mathrm{d}t \thinspace D|\bm\nabla V(\bm{x}_{\mspace{-1mu}t}, t)|^2
  \biggr\rangle + \beta \bigg\langle
    \int_0^{\mspace{-1mu}\tau} \negthickspace \mathrm{d}t \thinspace D \bm\nabla V(\bm{x}_{\mspace{-1mu}t}, t) \cdot \bm\nabla U(\bm{x}_{\mspace{-1mu}t}, t)
  \biggr\rangle
  \nonumber \\
  &= \biggl\langle
    \int_0^{\mspace{-1mu}\tau} \negthickspace \mathrm{d}t \thinspace D|\bm\nabla V(\bm{x}_{\mspace{-1mu}t}, t)|^2
  \biggr\rangle - \beta \bigg\langle
    \negthinspace\mathord{-}\int_0^{\mspace{-1mu}\tau} \negthickspace \mathrm{d}\bm{x}_{\mspace{-1mu}t} \circ \bm\nabla U(\bm{x}_{\mspace{-1mu}t}, t)
  \biggr\rangle
  \label{seq:StratonovichChainRuleStochastic2} \\
  &\equiv \biggl\langle
    \int_0^{\mspace{-1mu}\tau} \negthickspace \mathrm{d}t \thinspace D|\bm\nabla V(\bm{x}_{\mspace{-1mu}t}, t)|^2
  \biggr\rangle - \beta \langle \mathcal{Q}_\tau \rangle
  \label{seq:StratonovichChainRuleStochastic3}
\end{align}
\end{subequations}
where we used \cref{seq:StratonovichIdentity} in the first equality [\cref{seq:StratonovichChainRuleStochastic1}] and again in the third equality [\cref{seq:StratonovichChainRuleStochastic2}], and where we inserted the definition of the \emph{heat}, introduced in \cref{eq:PathHeat} of the main text, into the last equality [\cref{seq:StratonovichChainRuleStochastic3}].
Summing the results in \cref{seq:StratonovichChainRuleDeterministic,seq:StratonovichChainRuleStochastic} finally yields \cref{eq:Entropy} of the main text.

\section{Loss function for the nonadiabatic force}
\label{sec:LossDerivation}

In this section, we show that the loss function \( \ell(\bm{\theta}) \), introduced in \cref{eq:Loss} of the main text to train the nonadiabatic force ansatz \( \bm\nabla V_{\!\bm{\theta}} \), is equivalent to the mean-squared-error (MSE) loss function
\begin{align}
  \ell_\mathrm{MSE}(\bm{\theta})
  &\equiv
  \biggl\langle \int_0^{\mspace{-1mu}\tau} \negthinspace \mathrm{d}t \thinspace
    D| \bm\nabla V_{\!\bm{\theta}}(\bm{x}_{\mspace{-1mu}t}, t) - \bm\nabla V(\bm{x}_{\mspace{-1mu}t}, t) |^2
  \biggr\rangle
  \nonumber \\
  &=
  \biggl\langle \int_0^{\mspace{-1mu}\tau} \negthinspace \mathrm{d}t \thinspace
    D| \bm\nabla V_{\!\bm{\theta}}(\bm{x}_{\mspace{-1mu}t}, t)|^2 - 2D \bm\nabla V(\bm{x}_{\mspace{-1mu}t}, t) \cdot \bm\nabla V_{\!\bm{\theta}}(\bm{x}_{\mspace{-1mu}t}, t)
  \biggr\rangle + \text{const.}
  \label{seq:Loss}
\end{align}
where \( \bm\nabla V \) is the exact nonadiabatic force introduced in \cref{eq:FokkerPlanck} of the main text.
To demonstrate equivalence between \( \ell(\bm{\theta}) \) and \( \ell_\mathrm{MSE}(\bm{\theta}) \), it is sufficient to show that the cross-terms in both losses are identical; this amounts to proving the equality in \cref{seq:StratonovichIdentity}.
We apply the \emph{Stratonovich-to-It\={o} conversion} (a corollary of \emph{It\={o}'s lemma}; see, e.g., Lemma~3.2 from Ref.~\onlinecite{Pavliotis2014Book}) to the integrand in the left-hand side of \cref{seq:StratonovichIdentity}, obtaining
\begin{equation}\label{seq:StratonovichToIto}
  \int_0^{\mspace{-1mu}\tau} \negthinspace \mathrm{d}\bm{x}_{\mspace{-1mu}t}
  \circ \bm{f}(\bm{x}_{\mspace{-1mu}t}, t)
  =
  \int_0^{\mspace{-1mu}\tau} \negthinspace \mathrm{d}\bm{x}_{\mspace{-1mu}t}
  \cdot \bm{f}(\bm{x}_{\mspace{-1mu}t}, t)
  +
  \int_0^{\mspace{-1mu}\tau} \negthinspace \mathrm{d}t \thinspace
  \bm\nabla \cdot [D \bm{f}(\bm{x}_{\mspace{-1mu}t}, t)]
\end{equation}
where the first term on the right-hand side is an It\={o} stochastic integral against the process specified by \cref{eq:ForwardProcess} of the main text.
Using this equation, we rewrite the right-hand side of \cref{seq:StratonovichIdentity} as
\begin{align}
  \biggl\langle
    \int_0^{\mspace{-1mu}\tau} \negthinspace \mathrm{d}\bm{x}_{\mspace{-1mu}t}
    \cdot \bm{f}(\bm{x}_{\mspace{-1mu}t}, t)
  \biggr\rangle
  &=
  \biggl\langle
    \negthinspace\mathord{-}\int_0^{\mspace{-1mu}\tau} \negthinspace \mathrm{d}t \thinspace
    D\beta \bm\nabla U(\bm{x}_{\mspace{-1mu}t}, t) \cdot \bm{f}(\bm{x}_{\mspace{-1mu}t}, t)
  \biggr\rangle
  + 
  \biggl\langle
    \int_0^{\mspace{-1mu}\tau} \negthinspace
    \sqrt{2D} \thinspace \mathrm{d}\bm{w}_{\mspace{-1mu}t} \cdot \bm{f}(\bm{x}_{\mspace{-1mu}t}, t)
  \biggr\rangle
  \nonumber \\
  &\equiv
  \biggl\langle
    \int_0^{\mspace{-1mu}\tau} \negthinspace \mathrm{d}t \thinspace
    D \bm\nabla \ln \pi \bl( \bm{x}_{\mspace{-1mu}t}, \lambda(t) \br) \cdot \bm{f}(\bm{x}_{\mspace{-1mu}t}, t)
  \biggr\rangle
  \label{seq:StratonovichToItoStochastic}
\end{align}
Here, the second term in the right-hand side of the first equality vanishes due to the non-anticipative property of \( \{\bm{x}_{\mspace{-1mu}t}\}_{t=\mspace{1mu}0}^\tau \) and the zero-mean property of the Wiener process \( \{\bm{w}_{\mspace{-1mu}t}\}_{t=\mspace{1mu}0}^\tau \).
The second term on the right-hand side of \cref{seq:StratonovichToIto} can be brought to a similar form,
\begin{align}
  \biggl\langle
    \int_0^{\mspace{-1mu}\tau} \negthinspace \mathrm{d}t \thinspace
    \bm\nabla \cdot [D \bm{f}(\bm{x}_{\mspace{-1mu}t}, t)]
  \biggr\rangle
  &=
  \int_0^{\mspace{-1mu}\tau} \negthinspace \mathrm{d}t
  \int \mathrm{d}\bm{x} \thinspace
  \rho(\bm{x}, t) \thinspace \bm\nabla \cdot [D \bm{f}(\bm{x}, t)]
  \nonumber \\ &=
  \int_0^{\mspace{-1mu}\tau} \negthinspace \mathrm{d}t
  \thinspace \biggl[
  \mathord{-}\int \mathrm{d}\bm{x} \thinspace
  D \bm\nabla \rho(\bm{x}, t) \thinspace \cdot \bm{f}(\bm{x}, t) \biggr]
  \nonumber \\
  &=
  \biggl\langle
    \negthinspace\mathord{-}\int_0^{\mspace{-1mu}\tau} \negthinspace \mathrm{d}t \thinspace
    D \bm\nabla \ln \rho(\bm{x}_{\mspace{-1mu}t}, t) \cdot \bm{f}(\bm{x}_{\mspace{-1mu}t}, t)
  \biggr\rangle
  \label{seq:StratonovichToItoDeterministic}
\end{align}
where the second equality is obtained upon integrating by parts in the configurational variable while assuming vanishing surface terms.
Substituting \cref{seq:StratonovichToItoStochastic,seq:StratonovichToItoDeterministic} into \cref{seq:StratonovichToIto} finally yields
\begin{align}
  \biggl\langle
    \int_0^{\mspace{-1mu}\tau} \negthinspace \mathrm{d}\bm{x}_{\mspace{-1mu}t}
    \circ \bm{f}(\bm{x}_{\mspace{-1mu}t}, t)
  \biggr\rangle
  &=
  \biggl\langle
    \negthinspace\mathord{-}\int_0^{\mspace{-1mu}\tau} \negthinspace \mathrm{d}t \thinspace
     D \bm\nabla \ln \frac{\rho (\bm{x}_{\mspace{-1mu}t}, t)}{\pi \bl( \bm{x}_{\mspace{-1mu}t}, \lambda(t) \br)} \cdot \bm{f}(\bm{x}_{\mspace{-1mu}t}, t)
  \biggr\rangle
  \nonumber \\
  &\equiv
  \biggl\langle
    \int_0^{\mspace{-1mu}\tau} \negthinspace \mathrm{d}t \thinspace
     D \bm\nabla V(\bm{x}_{\mspace{-1mu}t}, t) \cdot \bm{f}(\bm{x}_{\mspace{-1mu}t}, t)
  \biggr\rangle
  \label{seq:StratonovichIdentityProof}
\end{align}
where \cref{seq:StratonovichIdentity} is recovered in the second equality per the definition of the nonadiabatic potential.

\section{Relation between the forward and backward processes}
\label{sec:PathRelativeDensityDerivation}


In this section, we derive the likelihood ratio in \cref{eq:PathRelativeDensityBackward} between the forward and backward processes specified by \cref{eq:ForwardProcess} and \cref{eq:BackwardProcess} of the main text.
We split our work into three subsections.
In \cref{ssec:Section_IV_A}, we identify the time-reversed process that the backward process seeks to approximate.
In \cref{ssec:Section_IV_B}, we provide a detailed derivation of the likelihood ratio for a backward process with a general nonadiabatic potential ansatz.
In \cref{ssec:Section_IV_C}, we show that the likelihood ratio reduces to unity for almost all configurational trajectories when the backward process is endowed with an optimal nonadiabatic potential.

\subsection{The forward process and its time-reversal}
\label{ssec:Section_IV_A}

In this subsection, we re-introduce the forward process alongside its time-reversal, and we derive relations between the two that clarify the role of the backward process as established in later subsections.
At times \( t \in [0, \tau] \), the forward process and its time-reversal respectively obey
\begin{equation}\label{seq:Processes}
  \mathrm{d}\bm{x}_{\mspace{-1mu}t} =
  -\beta D \bm\nabla U \bl( \bm{x}_{\mspace{-1mu}t}, \lambda(t) \br) \thinspace \mathrm{d}t
  + \sqrt{2D} \medspace \mathrm{d}\bm{w}_{\mspace{-1mu}t}
  \enspace\mathrm{and}\enspace
  \mathrm{d}\bm{x}_{\mspace{-1mu}t} =
  \beta D \bm\nabla U \bl( \bm{x}_{\mspace{-1mu}t}, \lambda(\tau\mspace{-1mu}\smallminus\mspace{-1mu}t) \br) \thinspace \mathrm{d}t
  + \sqrt{2D} \thinspace\mathord{\odot}\thinspace \mathrm{d}\overline{\bm{w}}_t
\end{equation}
with initial configurations \( \bm{x}_{\mspace{-1mu}0} \) respectively drawn from the distributions with densities \( \pi \bl( \,\cdot\,, \lambda(0) \br) \) and \( \rho( \,\cdot\,, \tau ) \).
The symbol \( \odot \) indicates that the backward process should be interpreted in the \emph{anti-It\={o}} sense~\cite[Section~2.7]{Kunita2019Book}, and \( \{\overline{\bm{w}}_t\}_{t=\mspace{1mu}0}^\tau \equiv \{\bm{w}_{\mspace{-1mu}t} - \bm{w}_{\mspace{-1mu}\tau}\}_{t=\mspace{1mu}0}^\tau \) is a time-reversed Wiener process.
The forward process and its time-reversal each induce a probability measure, respectively denoted by \( \mathbb{P}_{\mspace{-5mu}\tau}^{\rightarrow} \) and \( \mathbb{P}_{\mspace{-5mu}\tau}^{\leftarrow} \), on the space of configurational trajectories.

By the Girsanov theorem~\cite[Theorem~8.6.6]{Oksendal2003Book}, the forward process and its time-reversal each have a relative density with respect to a corresponding \emph{driftless} process (i.e., a process with \(U \equiv 0\)) endowed with the same initial condition.
Along a trajectory \( \bm{X} \medspace\mathord{\equiv}\medspace \{\bm{x}_{\mspace{-1mu}t}\}_{t=\mspace{1mu}0}^\tau \), the density of the forward process relative to its driftless counterpart is
\begin{equation}\label{seq:DriftlessGirsanovForward}
  \frac
  {\hfill\mathrm{d}\mathbb{P}_{\mspace{-5mu}\tau}^{\rightarrow}
  [\bm{X}]}
  {\hfill\mathrm{d}\mathbb{P}_{\mspace{-5mu}\tau,0}^{\rightarrow}
  [\bm{X}]}
  = \exp \biggl[
    \mathord{-}\frac{1}{4} \negthinspace \int_0^{\mspace{-1mu}\tau} \negthinspace
    D|\beta \bm\nabla U \bl( \bm{x}_{\mspace{-1mu}t}, \lambda(t) \br)|^2 \thinspace \mathrm{d}t
    + 2\beta \bm\nabla U \bl( \bm{x}_{\mspace{-1mu}t}, \lambda(t) \br) \cdot \mathrm{d}\bm{x}_{\mspace{-1mu}t}
  \biggr]
\end{equation}
and, for the time-reversed process, we have
\begin{equation}\label{seq:DriftlessGirsanovBackward}
  \frac
  {\hfill\mathrm{d}\mathbb{P}_{\mspace{-5mu}\tau}^{\leftarrow}
  [\bm{X}]}
  {\hfill\mathrm{d}\mathbb{P}_{\mspace{-5mu}\tau,0}^{\leftarrow}
  [\bm{X}]}
  = \exp \biggl[
    \mathord{-}\frac{1}{4} \negthinspace \int_0^{\mspace{-1mu}\tau} \negthinspace
    D|\beta \bm\nabla U \bl( \bm{x}_{\mspace{-1mu}t}, \lambda(\tau\mspace{-1mu}\smallminus\mspace{-1mu}t) \br)|^2 \thinspace \mathrm{d}t
    - 2\beta \bm\nabla U \bl( \bm{x}_{\mspace{-1mu}t}, \lambda(\tau\mspace{-1mu}\smallminus\mspace{-1mu}t) \br) \odot \mathrm{d}\bm{x}_{\mspace{-1mu}t}
  \biggr]
\end{equation}
where \( \mathbb{P}_{\mspace{-5mu}\tau,0}^{\rightarrow} \) and \( \mathbb{P}_{\mspace{-5mu}\tau,0}^{\leftarrow} \) denote the probability measures induced by the respective driftless processes.
Furthermore, the driftless processes are reversible with respect to one another, in the sense that the average of a test observable \( \mathcal{O}_\tau[\bm{X}] \) satisfies
\begin{equation}\label{seq:TimeReversalSymmetry}
  \bigl\langle
    \mathcal{O}_\tau[\bm{X}]
  \bigl\rangle_{\mathbb{P}_{\mspace{-5mu}\tau,0}^{\rightarrow}}
  \thickspace\equiv
  \biggl\langle
    \negthinspace \mathcal{O}_\tau[\mathsf{R} \bm{X}]
    \thinspace
    \frac
    {\hfill\mathrm{d}\mathbb{P}_{\mspace{-5mu}\tau,0}^{\rightarrow}
    [\mathsf{R} \bm{X}]}
    {\hfill\mathrm{d}\mathbb{P}_{\mspace{-5mu}\tau,0}^{\leftarrow}
    [\bm{X}]}
  \biggr\rangle_{\negthinspace\mathbb{P}_{\mspace{-5mu}\tau,0}^{\leftarrow}}
  \enspace\mathrm{and}\quad
  \bigl\langle
    \mathcal{O}_\tau[\bm{X}]
  \bigl\rangle_{\mathbb{P}_{\mspace{-5mu}\tau,0}^{\leftarrow}}
  \thickspace\equiv
  \biggl\langle
    \negthinspace \mathcal{O}_\tau[\mathsf{R} \bm{X}]
    \thinspace
    \frac
    {\hfill\mathrm{d}\mathbb{P}_{\mspace{-5mu}\tau,0}^{\leftarrow}
    [\mathsf{R} \bm{X}]}
    {\hfill\mathrm{d}\mathbb{P}_{\mspace{-5mu}\tau,0}^{\rightarrow}
    [\bm{X}]}
  \biggr\rangle_{\negthinspace\mathbb{P}_{\mspace{-5mu}\tau,0}^{\rightarrow}}
\end{equation}
where the subscript of each angled bracket indicates the process under which the average is taken, \( \mathsf{R} \) is an involution that reverses a trajectory in time as \( \mathsf{R} \bm{X} \medspace\mathord{\equiv}\medspace \{ \bm{x}_{\mspace{-1mu}\tau-t} \}_{t=\mspace{1mu}0}^\tau \), and we introduced the likelihood ratios
\begin{equation}\label{seq:DriftlessForwardBackwardLikelihoodRatio}
    \frac
    {\hfill\mathrm{d}\mathbb{P}_{\mspace{-5mu}\tau,0}^{\rightarrow}
    [\mathsf{R} \bm{X}]}
    {\hfill\mathrm{d}\mathbb{P}_{\mspace{-5mu}\tau,0}^{\leftarrow}
    [\bm{X}]}
    \equiv
    \frac
    {\pi \bl( \bm{x}_{\mspace{-1mu}\tau}, \lambda(0) \br)}
    {\rho(\bm{x}_{\mspace{-1mu}0}, \tau)}
    \quad\mathrm{and}\quad
    \frac
    {\hfill\mathrm{d}\mathbb{P}_{\mspace{-5mu}\tau,0}^{\leftarrow}
    [\mathsf{R} \bm{X}]}
    {\hfill\mathrm{d}\mathbb{P}_{\mspace{-5mu}\tau,0}^{\rightarrow}
    [\bm{X}]}
    \equiv
    \frac
    {\rho(\bm{x}_{\mspace{-1mu}\tau}, \tau)}
    {\pi \bl( \bm{x}_{\mspace{-1mu}0}, \lambda(0) \br)}
\end{equation}
of a trajectory and its time-reversal along the driftless processes.

The time-reversal equivalence between driftless processes in \cref{seq:TimeReversalSymmetry} extends to the processes in \cref{seq:Processes} due to a time-reversal symmetry of their respective densities.
To introduce it, note that for almost every trajectory \( \{\bm{x}_{\mspace{-1mu}t}\}_{t=\mspace{1mu}0}^\tau \), the It\={o} integral of the vector-valued map \( t \mapsto \bm{f}(\bm{x}_{\mspace{-1mu}t}, t) \) transforms under time-reversal of the trajectory according to the rule
\begin{equation}\label{seq:TimeReversalRule}
  \int_0^{\mspace{-1mu}\tau} \negthinspace \mathrm{d}\bm{x}_{\mspace{-1mu}t} \cdot \bm{f}(\bm{x}_{\mspace{-1mu}t}, t)
  \medspace\xrightarrow{\{ \bm{x}_{\mspace{-1mu}t} \}_{t=\mspace{1mu}0}^\tau \thinspace\rightarrow\thinspace \{ \bm{x}_{\mspace{-1mu}\tau-t} \}_{t=\mspace{1mu}0}^\tau}
  \medspace\mathord{-}\negthinspace \int_0^{\mspace{-1mu}\tau} \negthinspace \mathrm{d}\bm{x}_{\mspace{-1mu}t} \odot \bm{f}(\bm{x}_{\mspace{-1mu}t}, \tau\mspace{-1mu}\smallminus\mspace{-1mu}t)
\end{equation}
Applying this rule to \cref{seq:DriftlessGirsanovForward,seq:DriftlessGirsanovBackward} reveals the time-reversal symmetries
\begin{equation}\label{seq:DriftlessGirsanovTimeInvolution}
  \frac
  {\hfill\mathrm{d}\mathbb{P}_{\mspace{-5mu}\tau}^{\leftarrow}
  [\mathsf{R} \bm{X}]}
  {\hfill\mathrm{d}\mathbb{P}_{\mspace{-5mu}\tau,0}^{\leftarrow}
  [\mathsf{R} \bm{X}]}
  = \frac
  {\hfill\mathrm{d}\mathbb{P}_{\mspace{-5mu}\tau}^{\rightarrow}
  [\bm{X}]}
  {\hfill\mathrm{d}\mathbb{P}_{\mspace{-5mu}\tau,0}^{\rightarrow}
  [\bm{X}]}
  \quad\mathrm{and}\quad
  \frac
  {\hfill\mathrm{d}\mathbb{P}_{\mspace{-5mu}\tau}^{\rightarrow}
  [\mathsf{R} \bm{X}]}
  {\hfill\mathrm{d}\mathbb{P}_{\mspace{-5mu}\tau,0}^{\rightarrow}
  [\mathsf{R} \bm{X}]}
  = \frac
  {\hfill\mathrm{d}\mathbb{P}_{\mspace{-5mu}\tau}^{\leftarrow}
  [\bm{X}]} 
  {\hfill\mathrm{d}\mathbb{P}_{\mspace{-5mu}\tau,0}^{\leftarrow}
  [\bm{X}]}
\end{equation}
which allow converting between forward and time-reversed processes when evaluating expectation values.
Indeed, the expectation value of a test observable \( \mathcal{O}_\tau[\bm{X}] \) along the forward process is
\begin{subequations}\label{seq:ForwardBackwardDerivation}
\begin{align}
  \bigl\langle
    \mathcal{O}_\tau[\bm{X}]
  \bigr\rangle_{\mathbb{P}_{\mspace{-5mu}\tau}^{\rightarrow}}
  &=
  \biggl\langle
    \negthinspace \mathcal{O}_\tau[\bm{X}]
    \thinspace \frac
    {\hfill\mathrm{d}\mathbb{P}_{\mspace{-5mu}\tau}^{\rightarrow}
    [\bm{X}]}
    {\hfill\mathrm{d}\mathbb{P}_{\mspace{-5mu}\tau,0}^{\rightarrow}
    [\bm{X}]}
  \biggr\rangle_{\negthinspace\mathbb{P}_{\mspace{-5mu}\tau,0}^{\rightarrow}}
  \nonumber \\ &\equiv
  \biggl\langle
    \negthinspace \mathcal{O}_\tau[\mathsf{R} \bm{X}]
    \thinspace \frac
    {\hfill\mathrm{d}\mathbb{P}_{\mspace{-5mu}\tau}^{\rightarrow}
    [\mathsf{R}\bm{X}]}
    {\hfill\mathrm{d}\mathbb{P}_{\mspace{-5mu}\tau,0}^{\rightarrow}
    [\mathsf{R}\bm{X}]}
    \thinspace \frac
    {\hfill\mathrm{d}\mathbb{P}_{\mspace{-5mu}\tau,0}^{\leftarrow}
    [\mathsf{R} \bm{X}]}
    {\hfill\mathrm{d}\mathbb{P}_{\mspace{-5mu}\tau,0}^{\rightarrow}
    [\bm{X}]}
  \biggr\rangle_{\negthinspace\mathbb{P}_{\mspace{-5mu}\tau,0}^{\leftarrow}}
  \label{seq:ForwardBackwardDerivation1} \\ &=
  \biggl\langle
    \negthinspace \mathcal{O}_\tau[\mathsf{R} \bm{X}]
    \thinspace \frac
    {\hfill\mathrm{d}\mathbb{P}_{\mspace{-5mu}\tau}^{\rightarrow}
    [\mathsf{R}\bm{X}]}
    {\hfill\mathrm{d}\mathbb{P}_{\mspace{-5mu}\tau,0}^{\rightarrow}
    [\mathsf{R} \bm{X}]}
    \thinspace \frac
    {\hfill\mathrm{d}\mathbb{P}_{\mspace{-5mu}\tau,0}^{\leftarrow}
    [\mathsf{R} \bm{X}]}
    {\hfill\mathrm{d}\mathbb{P}_{\mspace{-5mu}\tau,0}^{\rightarrow}
    [\bm{X}]}
    \thinspace \frac
    {\hfill\mathrm{d}\mathbb{P}_{\mspace{-5mu}\tau,0}^{\leftarrow}
    [\bm{X}]}
    {\hfill\mathrm{d}\mathbb{P}_{\mspace{-5mu}\tau}^{\leftarrow}
    [\bm{X}]}
  \biggr\rangle_{\negthinspace\mathbb{P}_{\mspace{-5mu}\tau}^{\leftarrow}}
  \nonumber \\ &\equiv
  \biggl\langle
    \negthinspace \mathcal{O}_\tau[\mathsf{R} \bm{X}]
    \thinspace \frac
    {\hfill\mathrm{d}\mathbb{P}_{\mspace{-5mu}\tau}^{\rightarrow}
    [\mathsf{R} \bm{X}]}
    {\hfill\mathrm{d}\mathbb{P}_{\mspace{-5mu}\tau}^{\leftarrow}
    [\bm{X}]}
  \biggr\rangle_{\negthinspace\mathbb{P}_{\mspace{-5mu}\tau}^{\leftarrow}}
  \label{seq:ForwardBackwardDerivation2}
\end{align}
\end{subequations}
where the second equality [\cref{seq:ForwardBackwardDerivation1}] uses \cref{seq:TimeReversalSymmetry} and the last equality [\cref{seq:ForwardBackwardDerivation2}] defines the time-reversal likelihood ratio
\begin{equation}\label{seq:TimeReversalLikelihoodRatio1}
  \frac
  {\hfill\mathrm{d}\mathbb{P}_{\mspace{-5mu}\tau}^{\rightarrow}
  [\mathsf{R} \bm{X}]}
  {\hfill\mathrm{d}\mathbb{P}_{\mspace{-5mu}\tau}^{\leftarrow}
  [\bm{X}]}
  \equiv
  \frac
  {\hfill\mathrm{d}\mathbb{P}_{\mspace{-5mu}\tau}^{\rightarrow}
  [\mathsf{R} \bm{X}]}
  {\hfill\mathrm{d}\mathbb{P}_{\mspace{-5mu}\tau,0}^{\rightarrow}
  [\mathsf{R} \bm{X}]}
  \thinspace \frac
  {\hfill\mathrm{d}\mathbb{P}_{\mspace{-5mu}\tau,0}^{\rightarrow}
  [\mathsf{R} \bm{X}]}
  {\hfill\mathrm{d}\mathbb{P}_{\mspace{-5mu}\tau,0}^{\leftarrow}
  [\bm{X}]}
  \thinspace \frac
  {\hfill\mathrm{d}\mathbb{P}_{\mspace{-5mu}\tau,0}^{\leftarrow}
  [\bm{X}]}
  {\hfill\mathrm{d}\mathbb{P}_{\mspace{-5mu}\tau}^{\leftarrow}
  [\bm{X}]}
  =
  \frac{\pi \bl( \bm{x}_{\mspace{-1mu}\tau}, \lambda(0) \br)}{\rho(\bm{x}_{\mspace{-1mu}0}, \tau)}
\end{equation}
with \( \bm{X} \) a realization of the time-reversed process.
A similar derivation obtains
\begin{equation}\label{seq:TimeReversalLikelihoodRatio2}
  \frac
  {\hfill\mathrm{d}\mathbb{P}_{\mspace{-5mu}\tau}^{\leftarrow}
  [\mathsf{R} \bm{X}]}
  {\hfill\mathrm{d}\mathbb{P}_{\mspace{-5mu}\tau}^{\rightarrow}
  [\bm{X}]}
  \equiv
  \frac
  {\hfill\mathrm{d}\mathbb{P}_{\mspace{-5mu}\tau}^{\leftarrow}
  [\mathsf{R} \bm{X}]}
  {\hfill\mathrm{d}\mathbb{P}_{\mspace{-5mu}\tau,0}^{\leftarrow}
  [\mathsf{R} \bm{X}]}
  \thinspace \frac
  {\hfill\mathrm{d}\mathbb{P}_{\mspace{-5mu}\tau,0}^{\leftarrow}
  [\mathsf{R} \bm{X}]}
  {\hfill\mathrm{d}\mathbb{P}_{\mspace{-5mu}\tau,0}^{\rightarrow}
  [\bm{X}]}
  \thinspace \frac
  {\hfill\mathrm{d}\mathbb{P}_{\mspace{-5mu}\tau,0}^{\rightarrow}
  [\bm{X}]}
  {\hfill\mathrm{d}\mathbb{P}_{\mspace{-5mu}\tau}^{\rightarrow}
  [\bm{X}]}
  =
  \frac{\rho(\bm{x}_{\mspace{-1mu}\tau}, \tau)}{\pi \bl( \bm{x}_{\mspace{-1mu}0}, \lambda(0) \br)}
\end{equation}
with \( \bm{X} \) a realization of the forward process.

It is notable that the above likelihood ratios are independent of most details of a trajectory, essentially amounting to a reweighing of its endpoints.
The same property is shared by the likelihood of a trajectory along a given driftless process relative to its time-reversal along the same process, which takes the form
\begin{equation}\label{seq:TimeReversalLikelihoodRatio3}
  \frac
  {\hfill\mathrm{d}\mathbb{P}_{\mspace{-5mu}\tau,0}^{\rightarrow}
  [\mathsf{R} \bm{X}]}
  {\hfill\mathrm{d}\mathbb{P}_{\mspace{-5mu}\tau,0}^{\rightarrow}
  [\bm{X}]}
  =
  \frac
  {\pi \bl( \bm{x}_{\mspace{-1mu}\tau}, \lambda(0) \br)}
  {\pi \bl( \bm{x}_{\mspace{-1mu}0}, \lambda(0) \br)}
  \quad\mathrm{and}\quad
  \frac
  {\hfill\mathrm{d}\mathbb{P}_{\mspace{-5mu}\tau,0}^{\leftarrow}
  [\mathsf{R} \bm{X}]}
  {\hfill\mathrm{d}\mathbb{P}_{\mspace{-5mu}\tau,0}^{\leftarrow}
  [\bm{X}]}
  =
  \frac
  {\rho( \bm{x}_{\mspace{-1mu}\tau}, \tau)}
  {\rho( \bm{x}_{\mspace{-1mu}0}, \tau)}
\end{equation}
and will prove useful in the following subsections.

\subsection{The forward-backward likelihood ratio}
\label{ssec:Section_IV_B}

In this subsection, we re-introduce the backward process, originally introduced in \cref{eq:BackwardProcess} of the main text, as a parametric approximation of the reverse-time process from the previous subsection.
Recall that the backward process obeys
\begin{equation}\label{seq:BackwardProcess}
  \mathrm{d}\bm{x}_{\mspace{-1mu}t} =
  -D [ \beta \bm\nabla U \bl( \bm{x}_{\mspace{-1mu}t}, \lambda(\tau\mspace{-1mu}\smallminus\mspace{-1mu}t) \br) + 2 \bm\nabla V_{\!\bm\theta}(\bm{x}_{\mspace{-1mu}t}, \tau\mspace{-1mu}\smallminus\mspace{-1mu}t) ] \thinspace \mathrm{d}t
  + \sqrt{2D} \thinspace \mathrm{d}\bm{w}_{\mspace{-1mu}t}
  \enspace\mathrm{for}\enspace
  t \in [0, \tau]
\end{equation}
with initial configuration \( \bm{x}_{\mspace{-1mu}0} \) drawn from the distribution with density \( \rho(\,\cdot\,, \tau) \).
Like the forward process and its time-reversal, the backward process induces a path measure \( \mathbb{P}_{\mspace{-5mu}\tau,\bm\theta} \) on the space of configurational trajectories.
Given any such trajectory \( \bm{X} \medspace\mathord{\equiv}\medspace \{ \bm{x}_{\mspace{-1mu}t} \}_{t=\mspace{1mu}0}^\tau \), we seek an expression for its likelihood along the backward process relative to that of its time-reversal \( \mathsf{R} \bm{X} \medspace\mathord{\equiv}\medspace \{ \bm{x}_{\mspace{-1mu}\tau-t} \}_{t=\mspace{1mu}0}^\tau \) along the forward process.
This forward-backward likelihood ratio may be defined as
\begin{equation}\label{seq:ForwardBackwardLikelihoodRatioDefinition}
  \frac
  {\hfill\mathrm{d}\mathbb{P}_{\mspace{-5mu}\tau}
  [\{\bm{x}_{\mspace{-1mu}\tau-t}\}_{t=\mspace{1mu}0}^\tau]}
  {\hfill\mathrm{d}\mathbb{P}_{\mspace{-5mu}\tau,\bm\theta}
  [\{\bm{x}_{\mspace{-1mu}t}\}_{t=\mspace{1mu}0}^\tau]}
  \equiv
  \frac
  {\hfill\mathrm{d}\mathbb{P}_{\mspace{-5mu}\tau}^{\rightarrow}
  [\mathsf{R} \bm{X}]}
  {\hfill\mathrm{d}\mathbb{P}_{\mspace{-5mu}\tau,0}^{\rightarrow}
  [\bm{X}]}
  \thinspace
  \frac
  {\hfill\mathrm{d}\mathbb{P}_{\mspace{-5mu}\tau,0}^{\rightarrow}
  [ \bm{X}]}
  {\hfill\mathrm{d}\mathbb{P}_{\mspace{-5mu}\tau,\bm\theta}
  [ \bm{X}]}
\end{equation}
where the first factor on the right-hand side, which is the likelihood of the time-reversed trajectory \( \mathsf{R} \bm{X} \) under the forward process, reduces to
\begin{equation}\label{seq:TimeReversedLikelihood}
  \frac
  {\hfill\mathrm{d}\mathbb{P}_{\mspace{-5mu}\tau}^{\rightarrow}
  [\mathsf{R} \bm{X}]}
  {\hfill\mathrm{d}\mathbb{P}_{\mspace{-5mu}\tau,0}^{\rightarrow}
  [\bm{X}]}
  \equiv
  \frac
  {\hfill\mathrm{d}\mathbb{P}_{\mspace{-5mu}\tau,0}^{\rightarrow}
  [\mathsf{R} \bm{X}]}
  {\hfill\mathrm{d}\mathbb{P}_{\mspace{-5mu}\tau,0}^{\rightarrow}
  [\bm{X}]}
  \thinspace
  \frac
  {\hfill\mathrm{d}\mathbb{P}_{\mspace{-5mu}\tau}^{\rightarrow}
  [\mathsf{R} \bm{X}]}
  {\hfill\mathrm{d}\mathbb{P}_{\mspace{-5mu}\tau,0}^{\rightarrow}
  [\mathsf{R} \bm{X}]}
  =
  \frac
  {\pi \bl( \bm{x}_{\mspace{-1mu}\tau}, \lambda(0) \br)}
  {\pi \bl( \bm{x}_{\mspace{-1mu}0}, \lambda(0) \br)}
  \thinspace
  \frac
  {\hfill\mathrm{d}\mathbb{P}_{\mspace{-5mu}\tau}^{\leftarrow}
  [\bm{X}]}
  {\hfill\mathrm{d}\mathbb{P}_{\mspace{-5mu}\tau,0}^{\leftarrow}
  [\bm{X}]}
\end{equation}
by using \cref{seq:TimeReversalSymmetry,seq:TimeReversalLikelihoodRatio3}.
The second factor on the right-hand side of \cref{seq:ForwardBackwardLikelihoodRatioDefinition}, which corresponds to the likelihood of the trajectory \(\bm{X} \) along the backward process, is
\begin{align}
  \frac
  {\hfill\mathrm{d}\mathbb{P}_{\mspace{-5mu}\tau,0}^{\rightarrow}
  [\bm{X}]}
  {\hfill\mathrm{d}\mathbb{P}_{\mspace{-5mu}\tau,\bm\theta}
  [\bm{X}]}
  &\medspace\mathord{=}\medspace
  \frac
  {\pi\bl(\bm{x}_{\mspace{-1mu}0}, \lambda(0)\br)}
  {\pi\bl(\bm{x}_{\mspace{-1mu}\tau}, \lambda(0)\br)}
  \thinspace
  \frac
  {\pi\bl(\bm{x}_{\mspace{-1mu}\tau}, \lambda(0)\br)}
  {\rho(\bm{x}_{\mspace{-1mu}0}, \tau)}
  \nonumber \\
  &\medspace\phantom{=}\medspace\mathord{\times}\medspace
  \exp \biggl[
    \int_0^{\mspace{-1mu}\tau} \negthinspace
    D [ | \bm\nabla V_{\!\bm\theta}( \bm{x}_{\mspace{-1mu}t}, \tau\mspace{-1mu}\smallminus\mspace{-1mu}t) |^2 
    + \beta \bm\nabla U \bl( \bm{x}_{\mspace{-1mu}t}, \lambda(\tau\mspace{-1mu}\smallminus\mspace{-1mu}t) \br) \cdot \bm\nabla V_{\!\bm\theta}( \bm{x}_{\mspace{-1mu}t}, \tau\mspace{-1mu}\smallminus\mspace{-1mu}t) ] \thinspace \mathrm{d}t
    + \bm\nabla V_{\!\bm\theta}( \bm{x}_{\mspace{-1mu}t}, \tau\mspace{-1mu}\smallminus\mspace{-1mu}t) \cdot \mathrm{d}\bm{x}_{\mspace{-1mu}t}
  \biggr]
  \nonumber \\
  &\medspace\phantom{=}\medspace\mathord{\times}\medspace
  \exp \biggl[
    \frac{1}{4} \negthinspace \int_0^{\mspace{-1mu}\tau} \negthinspace
    D | \beta\bm\nabla U \bl( \bm{x}_{\mspace{-1mu}t}, \lambda(\tau\mspace{-1mu}\smallminus\mspace{-1mu}t) \br) |^2 \thinspace \mathrm{d}t +
    2 \beta\bm\nabla U \bl( \bm{x}_{\mspace{-1mu}t}, \lambda(\tau\mspace{-1mu}\smallminus\mspace{-1mu}t) \br) \cdot \mathrm{d}\bm{x}_{\mspace{-1mu}t}
  \biggr]
  \label{seq:BackwardLikelihood}
\end{align}
obtained by direct application of Girsanov's formula~\cite[Theorem~8.6.6]{Oksendal2003Book}.
The three factors on the right-hand side, one on each line, can be simplified with the following manipulations:

\noindent\paragraph*{First factor---}
Rewrite the second term of the first factor in \cref{seq:BackwardLikelihood} as
\begin{subequations}\label{seq:BoundaryTerm}
\begin{align}
  \frac{\pi\bl(\bm{x}_{\mspace{-1mu}\tau}, \lambda(0)\br)}{\rho(\bm{x}_{\mspace{-1mu}0}, \tau)}
  &=
  \frac{\pi\bl(\bm{x}_{\mspace{-1mu}0}, \lambda(\tau)\br)}{\rho(\bm{x}_{\mspace{-1mu}0}, \tau)}
  \thinspace
  \frac{\pi\bl(\bm{x}_{\mspace{-1mu}\tau}, \lambda(0)\br)}{\pi\bl(\bm{x}_{\mspace{-1mu}0}, \lambda(\tau)\br)}
  \nonumber \\ &=
  \exp\biggl[
    -
    \ln \frac
    {\rho(\bm{x}_{\mspace{-1mu}0}, \tau)}
    {\pi\bl(\bm{x}_{\mspace{-1mu}0}, \lambda(\tau)\br)}
    +
    \ln \frac
    {\pi \bl( \bm{x}_{\mspace{-1mu}\tau}, \lambda(0) \br)}
    {\pi \bl( \bm{x}_{\mspace{-1mu}0}, \lambda(\tau) \br)}
  \biggr]
  \nonumber \\ &\equiv
  \exp\biggl[
    V(\bm{x}_{\mspace{-1mu}0}, \tau)
    +
    \negthinspace \int_0^{\mspace{-1mu}\tau} \negthinspace
    \mathrm{d} \ln \pi \bl( \bm{x}_{\mspace{-1mu}t}, \lambda(\tau\mspace{-1mu}\smallminus\mspace{-1mu}t)\br)
  \biggr]
  \label{seq:BoundaryTerm1} \\ &=
  \exp\biggl[
    V(\bm{x}_{\mspace{-1mu}0}, \tau)
    - \beta \Delta F
    - \beta \negthinspace \int_0^{\mspace{-1mu}\tau} \negthinspace
    \bm\nabla U \bl( \bm{x}_{\mspace{-1mu}t}, \lambda(\tau\mspace{-1mu}\smallminus\mspace{-1mu}t)\br) \circ \mathrm{d}\bm{x}_{\mspace{-1mu}t} +
    \partial_t U \bl( \bm{x}_{\mspace{-1mu}t}, \lambda(\tau\mspace{-1mu}\smallminus\mspace{-1mu}t)\br) \thinspace \mathrm{d}t
  \biggr]
  \label{seq:BoundaryTerm2} \\ &\equiv
  \exp\bigl[
    -\beta \bigl( \overline{\mathcal{W}}_{\mspace{-2mu}\tau} + \Delta F \bigr) + V(\bm{x}_{\mspace{-1mu}0}, \tau) + \beta \overline{\mathcal{Q}}_\tau
  \bigr] 
  \label{seq:BoundaryTerm3}
\end{align}
\end{subequations}
Here, \cref{seq:BoundaryTerm1} introduced the definition of the nonadiabatic potential, \cref{seq:BoundaryTerm2} used the Stratonovich chain rule to expand the differential \( \mathrm{d} \ln \pi \bl( \bm{x}_{\mspace{-1mu}t}, \lambda(\tau\mspace{-1mu}\smallminus\mspace{-1mu}t) \br) \) along the trajectory \( \bm{X} \), and \cref{seq:BoundaryTerm3} introduced the reverse-time heat and work, which are
\begin{equation}\label{seq:BackwardHeatAndWork}
  \overline{\mathcal{Q}}_\tau \equiv 
  \mathord{-} \negthinspace \int_0^{\mspace{-1mu}\tau} \negthinspace
  \bm\nabla U \bl( \bm{x}_{\mspace{-1mu}t}, \lambda(\tau\mspace{-1mu}\smallminus\mspace{-1mu}t) \br) \circ \mathrm{d}\bm{x}_{\mspace{-1mu}t}
  \quad\mathrm{and}\quad
  \overline{\mathcal{W}}_{\mspace{-2mu}\tau} \equiv 
  \int_0^{\mspace{-1mu}\tau} \negthinspace
  \partial_t U \bl( \bm{x}_{\mspace{-1mu}t}, \lambda(\tau\mspace{-1mu}\smallminus\mspace{-1mu}t) \br) \thinspace \mathrm{d}t,
\end{equation}
respectively.

\noindent\paragraph*{Second factor---}
Using \cref{seq:BackwardProcess}, rewrite the second factor in \cref{seq:BackwardLikelihood} as \( \exp\bigl(-\overline{\mathcal{S}}_{\tau,\bm\theta}\bigr) \), where
\begin{equation}\label{seq:BackwardSurprisal}
  \overline{\mathcal{S}}_{\tau,\bm\theta}
  \equiv
  \int_0^{\mspace{-1mu}\tau} \negthinspace
  D | \bm\nabla V_{\!\bm\theta}( \bm{x}_{\mspace{-1mu}t}, \tau\mspace{-1mu}\smallminus\mspace{-1mu}t) |^2 \thinspace \mathrm{d}t -
  \sqrt{2D} \thinspace \bm\nabla V_{\!\bm\theta}( \bm{x}_{\mspace{-1mu}t}, \tau\mspace{-1mu}\smallminus\mspace{-1mu}t) \cdot \mathrm{d}\bm{w}_{\mspace{-1mu}t}
\end{equation}
is the reverse-time surprisal defined in \cref{eq:BackwardSurprisal} of the main text.

\noindent\paragraph*{Third factor---}
For almost every trajectory \( \{\bm{x}_{\mspace{-1mu}t}\}_{t=\mspace{1mu}0}^\tau \), a Stratonovich integral of the vector-valued map \( t \mapsto \bm{f}(\bm{x}_{\mspace{-1mu}t}, t) \) along the trajectory can be expressed as a sum of It\={o} and anti-It\={o} integrals using the relation
\begin{equation}\label{seq:StratonovichFromItoAndAntiIto}
  \int_0^{\mspace{-1mu}\tau} \negthinspace \bm{f}(\bm{x}_{\mspace{-1mu}t}, t) \circ \mathrm{d} \bm{x}_{\mspace{-1mu}t}
  = \frac{1}{2} \thinspace \biggl[
    \int_0^{\mspace{-1mu}\tau} \negthinspace \bm{f}(\bm{x}_{\mspace{-1mu}t}, t) \cdot \mathrm{d} \bm{x}_{\mspace{-1mu}t} +
    \int_0^{\mspace{-1mu}\tau} \negthinspace \bm{f}(\bm{x}_{\mspace{-1mu}t}, t) \odot \mathrm{d} \bm{x}_{\mspace{-1mu}t}
  \biggr]
\end{equation}
Applying this relation to the third factor in \cref{seq:BackwardLikelihood} yields
\begin{align}
  &\exp \biggl[
    \frac{1}{4} \negthinspace \int_0^{\mspace{-1mu}\tau} \negthinspace
    D | \beta\bm\nabla U \bl( \bm{x}_{\mspace{-1mu}t}, \lambda(\tau\mspace{-1mu}\smallminus\mspace{-1mu}t) \br) |^2 \thinspace \mathrm{d}t
    + 2 \beta\bm\nabla U \bl( \bm{x}_{\mspace{-1mu}t}, \lambda(\tau\mspace{-1mu}\smallminus\mspace{-1mu}t) \br) \cdot \mathrm{d}\bm{x}_{\mspace{-1mu}t}
  \biggr]
  \nonumber \\
  &\mspace{60mu}\mathord{=}
  \medspace\exp \biggl[
    \frac{1}{4} \negthinspace \int_0^{\mspace{-1mu}\tau} \negthinspace
    D | \beta\bm\nabla U \bl( \bm{x}_{\mspace{-1mu}t}, \lambda(\tau\mspace{-1mu}\smallminus\mspace{-1mu}t) \br) |^2 \thinspace \mathrm{d}t
    - 2 \beta\bm\nabla U \bl( \bm{x}_{\mspace{-1mu}t}, \lambda(\tau\mspace{-1mu}\smallminus\mspace{-1mu}t) \br) \odot \mathrm{d}\bm{x}_{\mspace{-1mu}t}
  \biggr]
  \nonumber \\
  &\mspace{60mu}\phantom{=}\thickspace\mathord{\times}
  \medspace\exp \biggl[
    \frac{\beta}{2} \negthinspace \int_0^{\mspace{-1mu}\tau} \negthinspace
    \bm\nabla U \bl( \bm{x}_{\mspace{-1mu}t}, \lambda(\tau\mspace{-1mu}\smallminus\mspace{-1mu}t) \br) \odot \mathrm{d}\bm{x}_{\mspace{-1mu}t} +
    \bm\nabla U \bl( \bm{x}_{\mspace{-1mu}t}, \lambda(\tau\mspace{-1mu}\smallminus\mspace{-1mu}t) \br) \cdot \mathrm{d}\bm{x}_{\mspace{-1mu}t}
  \biggr]
  \nonumber \\
  &\mspace{60mu}\mathord{=}\medspace
  \frac
  {\hfill\mathrm{d}\mathbb{P}_{\mspace{-5mu}\tau,0}^{\leftarrow}
  [\bm{X}]}
  {\hfill\mathrm{d}\mathbb{P}_{\mspace{-5mu}\tau}^{\leftarrow}
  [\bm{X}]}
  \nonumber \\
  &\mspace{60mu}\phantom{=}\thickspace\mathord{\times}
  \medspace\exp \bigl( -\beta \overline{\mathcal{Q}}_\tau \bigr)
  \label{seq:InteriorTerm}
\end{align}
where the last equality uses \cref{seq:DriftlessGirsanovBackward} and \cref{seq:StratonovichFromItoAndAntiIto} to introduce the reverse-time heat in \cref{seq:BackwardHeatAndWork}.

Substituting \cref{seq:BoundaryTerm,seq:BackwardSurprisal,seq:InteriorTerm} into \cref{seq:BackwardLikelihood}, we obtain
\begin{align}
  \frac
  {\hfill\mathrm{d}\mathbb{P}_{\mspace{-5mu}\tau,0}^{\rightarrow}
  [\bm{X}]}
  {\hfill\mathrm{d}\mathbb{P}_{\mspace{-5mu}\tau,\bm\theta}
  [\bm{X}]}
  &\medspace\mathord{=}\medspace
  \exp\bigl[
    - \beta \bigl( \overline{\mathcal{W}}_{\mspace{-2mu}\tau} + \Delta F \bigr)
    + V(\bm{x}_{\mspace{-1mu}0}, \tau) 
    - \overline{\mathcal{S}}_{\tau,\bm\theta} 
  \bigr]
  \thinspace
  \nonumber \\
  &\medspace\phantom{=}\medspace\mathord{\times}\medspace
  \frac
  {\pi \bl( \bm{x}_{\mspace{-1mu}0}, \lambda(0) \br)}
  {\pi \bl( \bm{x}_{\mspace{-1mu}\tau}, \lambda(0) \br)}
  \thinspace
  \frac
  {\hfill\mathrm{d}\mathbb{P}_{\mspace{-5mu}\tau,0}^{\leftarrow}
  [\bm{X}]}
  {\hfill\mathrm{d}\mathbb{P}_{\mspace{-5mu}\tau}^{\leftarrow}
  [\bm{X}]}
  \label{seq:BackwardLikelihoodFinal}
\end{align}
a simplified expression for the likelihood of a backward trajectory, which upon multiplying by \cref{seq:TimeReversedLikelihood} finally becomes
\begin{equation}\label{seq:ForwardBackwardLikelihoodRatio}
  \frac
  {\hfill\mathrm{d}\mathbb{P}_{\mspace{-5mu}\tau}
  [\{ \bm{x}_{\mspace{-1mu}\tau-t} \}_{t=\mspace{1mu}0}^\tau]}
  {\hfill\mathrm{d}\mathbb{P}_{\mspace{-5mu}\tau,\bm\theta}
  [\{ \bm{x}_{\mspace{-1mu}t} \}_{t=\mspace{1mu}0}^\tau]}
  =
  \exp\bigl[
    - \beta \bigl( \overline{\mathcal{W}}_{\mspace{-2mu}\tau} + \Delta F \bigr)
    + V(\bm{x}_{\mspace{-1mu}0}, \tau) 
    - \overline{\mathcal{S}}_{\tau,\bm\theta} 
  \bigr]
\end{equation}
the expression in \cref{eq:PathRelativeDensityBackward} of the main text.

The forward-backward likelihood ratio derived above must be understood in reference to the backward process.
An analogous derivation, which we omit for brevity, obtains
\begin{equation}\label{seq:BackwardForwardLikelihoodRatio}
  \frac
  {\hfill\mathrm{d}\mathbb{P}_{\mspace{-5mu}\tau,\bm\theta}[\{ \bm{x}_{\mspace{-1mu}\tau-t} \}_{t=\mspace{1mu}0}^\tau]}
  {\hfill\mathrm{d}\mathbb{P}_{\mspace{-5mu}\tau}[\{ \bm{x}_{\mspace{-1mu}t} \}_{t=\mspace{1mu}0}^\tau]}
  =
  \exp\bigl[
    - \beta \bigl( \mathcal{W}_{\mspace{-2mu}\tau} - \Delta F \bigr)
    - V(\bm{x}_{\mspace{-1mu}\tau}, \tau) 
    + \mathcal{S}_{\tau,\bm\theta} 
  \bigr]
\end{equation}
a backward-forward likelihood ratio that is defined in reference to the forward process, where
\begin{equation}\label{seq:ForwardHeatAndWork}
  \mathcal{Q}_\tau
  \equiv \mathord{-} \negthinspace \int_0^{\mspace{-1mu}\tau} \negthinspace
  \bm\nabla U \bl( \bm{x}_{\mspace{-1mu}t}, \lambda(t) \br)
  \thinspace\mathord{\circ}\thinspace \mathrm{d}\bm{x}_{\mspace{-1mu}t}
  \quad\mathrm{and}\quad
  \mathcal{W}_{\mspace{-2mu}\tau}
  \equiv \int_0^{\mspace{-1mu}\tau} \negthinspace
  \partial_t U \bl( \bm{x}_{\mspace{-1mu}t}, \lambda(t) \br)
  \thinspace \mathrm{d}t
\end{equation}
are the heat dissipated by, and the work done on, a system evolving along the trajectory \( \bm{X} \), and where
\begin{equation}\label{seq:ForwardSurprisal}
  \mathcal{S}_{\tau,\bm\theta}
  \equiv
  \int_0^{\mspace{-1mu}\tau} \negthinspace
  D | \bm\nabla V_{\!\bm\theta}( \bm{x}_{\mspace{-1mu}t}, t) |^2 \thinspace \mathrm{d}t -
  \sqrt{2D} \thinspace \bm\nabla V_{\!\bm\theta}( \bm{x}_{\mspace{-1mu}t}, t) \cdot \mathrm{d}\bm{w}_{\mspace{-1mu}t}
\end{equation}
is the surprisal along the forward process.
Though not introduced in the main text, \cref{seq:BackwardForwardLikelihoodRatio} is key to derive the forward free-energy bound in \cref{eq:ForwardFreeEnergyBound} of the main text, as seen in \cref{sec:FreeEnergyBoundsDerivation} of this document.

\subsection{The forward-backward likelihood ratio given an optimal nonadiabatic potential}
\label{ssec:Section_IV_C}

To see that the backward process can become an exact approximation of the time-reversed process, we evaluate the likelihood ratio derived in the previous subsection given an optimal nonadiabatic potential \( V(\bm{x}, t) \equiv V_{\!\bm\theta^\star}(\bm{x}, t) - \ln \langle \exp[V_{\!\bm\theta^\star}(\bm{x}_{\mspace{-1mu}t}, t)] \rangle_{\mathbb{P}_{\mspace{-5mu}\tau}^{\rightarrow}} \) that satisfies \cref{eq:OptimalNonadiabaticPotential} of the main text.
Define
\begin{equation}\label{seq:ExactSurprisal}
  \overline{\mathcal{S}}_\tau
  \equiv
  \mathord{-} \negthinspace \int_0^{\mspace{-1mu}\tau} \negthinspace
    D [
    |\bm\nabla V(\bm{x}_{\mspace{-1mu}t}, \tau\mspace{-1mu}\smallminus\mspace{-1mu}t)|^2
    +
    \bm\nabla U \bl( \bm{x}_{\mspace{-1mu}t}, \lambda(\tau\mspace{-1mu}\smallminus\mspace{-1mu}t) \br)
    \cdot \bm\nabla V(\bm{x}_{\mspace{-1mu}t}, \tau\mspace{-1mu}\smallminus\mspace{-1mu}t)
    ] \thinspace \mathrm{d}t
    +
    \bm\nabla V(\bm{x}_{\mspace{-1mu}t}, \tau\mspace{-1mu}\smallminus\mspace{-1mu}t) \cdot \mathrm{d}\bm{x}_{\mspace{-1mu}t}
\end{equation}
by substituting \( V_{\!\bm\theta}(\bm{x}, t) \rightsquigarrow V(\bm{x}, t) \) in the definition of \( \overline{\mathcal{S}}_{\tau,\bm\theta} \) in \cref{seq:BackwardSurprisal}, and by using \cref{seq:BackwardProcess} to rewrite the integral against the Wiener process.
We now show that, given (almost) any trajectory \( \{ \bm{x}_t \}_{t=\mspace{1mu}0}^\tau \), the remaining terms in the forward-backward likelihood ratio [\cref{seq:ForwardBackwardLikelihoodRatio}] reduce to \cref{seq:ExactSurprisal}, rendering the likelihood ratio unity for almost every trajectory.
Indeed,
\begin{align}
  V(\bm{x}_{\mspace{-1mu}0}, \tau)
  - \beta \bigl( \overline{\mathcal{W}}_{\mspace{-2mu}\tau} + \Delta F \bigr)
  &= V(\bm{x}_{\mspace{-1mu}0}, \tau) - V(\bm{x}_{\mspace{-1mu}\tau}, 0)
  - \beta \bigl( \overline{\mathcal{W}}_{\mspace{-2mu}\tau} + \Delta F \bigr)
  \nonumber \\[0.25em]
  &= \mathord{-} \negthinspace \int_0^{\mspace{-1mu}\tau} \negthinspace
    \partial_t V(\bm{x}_{\mspace{-1mu}t}, \tau\mspace{-1mu}\smallminus\mspace{-1mu}t) \thinspace \mathrm{d}t + 
    \bm\nabla V(\bm{x}_{\mspace{-1mu}t}, \tau\mspace{-1mu}\smallminus\mspace{-1mu}t) \thinspace\mathord{\circ}\thinspace \mathrm{d}\bm{x}_{\mspace{-1mu}t}
  - \beta \bigl( \overline{\mathcal{W}}_{\mspace{-2mu}\tau} + \Delta F \bigr)
  \nonumber \\
  &= \mathord{-} \negthinspace \int_0^{\mspace{-1mu}\tau} \negthinspace
    [\mathord{-}\partial_t \ln \rho(\bm{x}_{\mspace{-1mu}t}, \tau\mspace{-1mu}\smallminus\mspace{-1mu}t)] \thinspace \mathrm{d}t
    + \bm\nabla V(\bm{x}_{\mspace{-1mu}t}, \tau\mspace{-1mu}\smallminus\mspace{-1mu}t) \thinspace\mathord{\circ}\thinspace \mathrm{d}\bm{x}_{\mspace{-1mu}t}
  \label{seq:ForwardBackwardExactDerivation1}
\end{align}
where the first equality uses the boundary condition \( V(\,\cdot\,, t \smallequals 0) \equiv 0 \), the second equality uses the Stratonovich chain rule to expand the differential \( \mathrm{d} V(\bm{x}_{\mspace{-1mu}t}, \tau\mspace{-1mu}\smallminus\mspace{-1mu}t) = \mathrm{d} \ln \pi \bl( \bm{x}_{\mspace{-1mu}t}, \lambda(\tau\mspace{-1mu}\smallminus\mspace{-1mu}t) \br) - \mathrm{d} \ln \rho(\bm{x}_{\mspace{-1mu}t}, \tau\mspace{-1mu}\smallminus\mspace{-1mu}t) \) along the trajectory, and the final equality uses the definition of the reverse-time work \( \overline{\mathcal{W}}_{\mspace{-2mu}\tau} \) in \cref{seq:BackwardHeatAndWork}.
Now, note that the Fokker--Planck equation solved by \( \rho(\bm{x}, t) \), stated in \cref{eq:FokkerPlanck} of the main text, can be rewritten to describe the time evolution of the time-reversed log-density \( \ln \rho(\bm{x}, \tau\mspace{-1mu}\smallminus\mspace{-1mu}t) \) as
\begin{equation}\label{seq:FokkerPlanckLogDensity}
  -\partial_t \ln \rho(\bm{x}, \tau\mspace{-1mu}\smallminus\mspace{-1mu}t)
  = D [ 
    | \bm\nabla V(\bm{x}, \tau\mspace{-1mu}\smallminus\mspace{-1mu}t) |^2
    + \beta \bm\nabla U\bl(\bm{x}, \lambda(\tau\mspace{-1mu}\smallminus\mspace{-1mu}t) \br)
      \cdot \bm\nabla V(\bm{x}, \tau\mspace{-1mu}\smallminus\mspace{-1mu}t)
  ]
  - \bm\nabla \cdot [ D \bm\nabla V(\bm{x}, \tau\mspace{-1mu}\smallminus\mspace{-1mu}t) ]
\end{equation}
where substituting \( t \rightsquigarrow \tau\mspace{-1mu}\smallminus\mspace{-1mu}t \) introduced the negative sign on the left-hand side.
Inserting this equation into \cref{seq:ForwardBackwardExactDerivation1}, we get
\begin{align}
  V(\bm{x}_{\mspace{-1mu}0}, \tau)
  - \beta \bigl( \overline{\mathcal{W}}_{\mspace{-2mu}\tau} + \Delta F \bigr)
  &= \mathord{-} \negthinspace \int_0^{\mspace{-1mu}\tau} \negthinspace
    D [
      | \bm\nabla V(\bm{x}_{\mspace{-1mu}t}, \tau\mspace{-1mu}\smallminus\mspace{-1mu}t) |^2
      + \beta \bm\nabla U\bl(\bm{x}_{\mspace{-1mu}t}, \lambda(\tau\mspace{-1mu}\smallminus\mspace{-1mu}t) \br)
        \cdot \bm\nabla V(\bm{x}_{\mspace{-1mu}t}, \tau\mspace{-1mu}\smallminus\mspace{-1mu}t)
    ]
    \thinspace \mathrm{d}t
    \nonumber \\
    &\mspace{65mu}
    \mathord{+} \medspace \int_0^{\mspace{-1mu}\tau} \negthinspace
    \bm\nabla V(\bm{x}_{\mspace{-1mu}t}, \tau\mspace{-1mu}\smallminus\mspace{-1mu}t) \thinspace\mathord{\circ}\thinspace \mathrm{d}\bm{x}_{\mspace{-1mu}t}
    - \bm\nabla \cdot [ D \bm\nabla V(\bm{x}_{\mspace{-1mu}t}, \tau\mspace{-1mu}\smallminus\mspace{-1mu}t) ] \thinspace \mathrm{d}t
  \nonumber \\
  &= \mathord{-} \negthinspace \int_0^{\mspace{-1mu}\tau} \negthinspace
    D [
      | \bm\nabla V(\bm{x}_{\mspace{-1mu}t}, \tau\mspace{-1mu}\smallminus\mspace{-1mu}t) |^2
      + \beta \bm\nabla U\bl(\bm{x}_{\mspace{-1mu}t}, \lambda(\tau\mspace{-1mu}\smallminus\mspace{-1mu}t) \br)
        \cdot \bm\nabla V(\bm{x}_{\mspace{-1mu}t}, \tau\mspace{-1mu}\smallminus\mspace{-1mu}t)
    ]
    \thinspace \mathrm{d}t
    \nonumber \\
    &\mspace{65mu}
    \mathord{+} \medspace \int_0^{\mspace{-1mu}\tau} \negthinspace
    \bm\nabla V(\bm{x}_{\mspace{-1mu}t}, \tau\mspace{-1mu}\smallminus\mspace{-1mu}t) \thinspace\mathord{\cdot}\thinspace \mathrm{d}\bm{x}_{\mspace{-1mu}t}
  \nonumber \\
  &= \overline{\mathcal{S}}_\tau
  \label{seq:ForwardBackwardExactDerivation2}
\end{align}
where the second equality used \cref{seq:StratonovichToIto} to transform the Stratonovich integral into an It\={o} integral.

We have thus shown that, given an optimal nonadiabatic potential, the likelihood ratio in \cref{seq:ForwardBackwardLikelihoodRatio} reduces to unity for almost all realizations of the backward process.
Taking the likelihood ratio as a measure of alignment between the time-reversed process and the backward process, we deduce that the two processes are \emph{probabilistically equal}, meaning that the expectation value of any observable may be written as an average of either process without reweighing.

\section{Variational bounds on the free-energy difference}
\label{sec:FreeEnergyBoundsDerivation}

In the following two subsections, we derive the lower and upper bounds on the free-energy difference introduced in \cref{eq:BackwardFreeEnergyBound} and \cref{eq:ForwardFreeEnergyBound} of the main text, which we respectively dub the \emph{backward bound} and the \emph{forward bound}.

\subsection{The backward (lower) bound on \texorpdfstring{\(\Delta F\)}{ΔF}}

An argument to obtain \cref{eq:BackwardFreeEnergyBound} of the main text is outlined therein up to the appearance of the learned nonadiabatic potential, \( V_{\!\bm\theta}(\bm{x}, \tau) - \ln \langle \exp[ V_{\!\bm\theta}(\bm{x}_{\mspace{-1mu}\tau}, \tau) ] \rangle \), in place of its exact counterpart, \( V(\bm{x}, \tau) \).
To justify this, notice that if the learned nonadiabatic potential \( V_{\!\bm\theta}(\bm{x}, \tau) \) is bounded from below, its cumulant-shifted exponential may be expressed as a ratio of two probability densities,
\begin{equation}
  \frac
  {\exp[ V_{\!\bm\theta}(\bm{x}, \tau) ]}
  {\langle\exp[ V_{\!\bm\theta}(\bm{x}_{\mspace{-1mu}\tau}, \tau) ]\rangle}
  \equiv 
  \frac
  {\pi \bl( \bm{x}, \lambda(\tau) \br)}
  {\rho_{\bm\theta}(\bm{x}, \tau)},
\end{equation}
where both the equilibrium density \( \pi \bl( \bm{x}, \lambda(\tau) \br) \) and the learned nonequilibrium density \( \rho_{\bm\theta}( \bm{x}, \tau ) \) are positive.
Taking logarithms and averaging with respect to the initial distribution of the backward process on both sides yields
\begin{align}
  \langle V_{\!\bm\theta}(\bm{x}_{\mspace{-1mu}\tau}, \tau) \rangle - \ln \langle \exp[ V_{\!\bm\theta}(\bm{x}_{\mspace{-1mu}\tau}, \tau) ] \rangle
  &\equiv
  \int \negthinspace \mathrm{d}\bm{x} \thinspace 
  \rho(\bm{x}, \tau) \ln \frac
  {\pi \bl( \bm{x}, \lambda(\tau) \br)}
  {\rho_{\bm\theta}(\bm{x}, \tau)}
  \nonumber \\ &=
  \int \negthinspace \mathrm{d}\bm{x} \thinspace 
  \rho(\bm{x}, \tau) \ln \frac
  {\rho(\bm{x}, \tau)}
  {\rho_{\bm\theta}(\bm{x}, \tau)}
  +
  \int \negthinspace \mathrm{d}\bm{x} \thinspace 
  \rho(\bm{x}, \tau) \ln \frac
  {\pi \bl( \bm{x}, \lambda(\tau) \br)}
  {\rho(\bm{x}, \tau)}
  \nonumber \\ &\ge
  -\int \negthinspace \mathrm{d}\bm{x} \thinspace 
  \rho(\bm{x}, \tau) \ln \frac
  {\rho(\bm{x}, \tau)}
  {\pi \bl( \bm{x}, \lambda(\tau) \br)}
  \nonumber \\ &=
  \langle V(\bm{x}_{\mspace{-1mu}\tau}, \tau) \rangle
  \label{seq:NonadiabaticPotentialInequality}
\end{align}
an inequality between the learned and exact averaged nonadiabatic potentials that follows from the non-negativity of Kullback--Leibler divergence.
After substituting \cref{eq:PathRelativeDensityBackward} into \cref{eq:PathKullbackLeiblerBackward}, the inequality in \cref{seq:NonadiabaticPotentialInequality} may be invoked to finally obtain \cref{eq:BackwardFreeEnergyBound} of the main text.

\subsection{The forward (upper) bound on \texorpdfstring{\(\Delta F\)}{ΔF}}

In this subsection, we derive \cref{eq:ForwardFreeEnergyBound} of the main text.
Because the time-reversal operator \( \mathsf{R} \) is a involution on the space of configurational trajectories, the pushforward of \( \mathbb{P}_{\mspace{-5mu}\tau,\bm\theta} \) by \( \mathsf{R} \), denoted \( \mathsf{R}\sharp\mathbb{P}_{\mspace{-5mu}\tau,\bm\theta} \), remains a probability measure and thus satisfies \( \langle 1 \rangle_{\mathsf{R}\sharp\mathbb{P}_{\mspace{-5mu}\tau,\bm\theta}} = 1 \).
Taking logarithms in this equality and using the likelihood ratio in \cref{seq:BackwardForwardLikelihoodRatio} to reweigh onto the forward process, we have
\begin{subequations}\label{seq:ForwardFreeEnergyBoundDerivation}
\begin{align}
  0 = \ln \thinspace \langle 1 \rangle_{\mathsf{R}\sharp\mathbb{P}_{\mspace{-5mu}\tau,\bm\theta}}
  &=
  \ln \thinspace \biggl\langle
    \frac
    {\hfill\mathrm{d}\mathbb{P}_{\mspace{-5mu}\tau,\bm\theta}
    [\mathsf{R} \bm{X}]}
    {\hfill\mathrm{d}\mathbb{P}_{\mspace{-5mu}\tau}^{\rightarrow}
    [\bm{X}]}
  \biggr\rangle_{\negthinspace\mathbb{P}_{\mspace{-5mu}\tau}^{\rightarrow}}
  \nonumber
  \\[0.5em] &=
  \ln \thinspace \bigl\langle
    \exp \bigl[
      - \beta \bigl( \mathcal{W}_{\mspace{-2mu}\tau} - \Delta F \bigr)
      - V(\bm{x}_{\mspace{-1mu}\tau}, \tau)
      + \mathcal{S}_{\tau,\bm\theta}
    \bigr]
  \bigr\rangle_{\mathbb{P}_{\mspace{-5mu}\tau}^{\rightarrow}}
  \label{seq:ForwardFreeEnergyBoundDerivation1}
  \\[0.5em] &\ge
  \bigl\langle
    - \beta \bigl( \mathcal{W}_{\mspace{-2mu}\tau} - \Delta F \bigr)
    - V(\bm{x}_{\mspace{-1mu}\tau}, \tau)
    + \mathcal{S}_{\tau,\bm\theta}
  \bigr\rangle_{\mathbb{P}_{\mspace{-5mu}\tau}^{\rightarrow}}
  \label{seq:ForwardFreeEnergyBoundDerivation2}
  \\[0.5em] &=
  \bigl\langle
    -\beta \bigl( \mathcal{W}_{\mspace{-2mu}\tau} - \Delta F \bigr)
  \bigr\rangle_{\mathbb{P}_{\mspace{-5mu}\tau}^{\rightarrow}}
  +
  \biggl\langle
    \int_0^{\mspace{-1mu}\tau} \negthinspace \mathrm{d}t \thinspace D | \bm\nabla V_{\!\bm\theta}(\bm{x}_{\mspace{-1mu}t}, t) |^2 - V(\bm{x}_{\mspace{-1mu}\tau}, \tau)
  \biggr\rangle_{\negthinspace\mathbb{P}_{\mspace{-5mu}\tau}^{\rightarrow}}
  \label{seq:ForwardFreeEnergyBoundDerivation3}
  \\ &\ge
  \bigl\langle
    -\beta \bigl( \mathcal{W}_{\mspace{-2mu}\tau} - \Delta F \bigr)
  \bigr\rangle_{\mathbb{P}_{\mspace{-5mu}\tau}^{\rightarrow}}
  +
  \biggl\langle
    \int_0^{\mspace{-1mu}\tau} \negthinspace \mathrm{d}t \thinspace D | \bm\nabla V_{\!\bm\theta}(\bm{x}_{\mspace{-1mu}t}, t) |^2
  \biggr\rangle_{\mathbb{P}_{\mspace{-5mu}\tau}^{\rightarrow}}
  \nonumber
  \\ &\medspace\phantom{=}\mathbin{-}
  \bigl[
    \langle V_{\!\bm\theta}(\bm{x}_{\mspace{-1mu}\tau}, \tau) \rangle_{\mathbb{P}_{\mspace{-5mu}\tau}^{\rightarrow}} -
    \ln \langle \exp[V_{\!\bm\theta}(\bm{x}_{\mspace{-1mu}\tau}, \tau)] \rangle_{\mathbb{P}_{\mspace{-5mu}\tau}^{\rightarrow}}
  \bigr]
  \label{seq:ForwardFreeEnergyBoundDerivation4}
\end{align}
\end{subequations}
Here, \cref{seq:ForwardFreeEnergyBoundDerivation1} uses \cref{seq:TimeReversalLikelihoodRatio2}, \cref{seq:ForwardFreeEnergyBoundDerivation2} uses Jensen's inequality, \cref{seq:ForwardFreeEnergyBoundDerivation3} expands the forward surprisal \( \mathcal{S}_{\tau,\bm\theta} \) from \cref{seq:ForwardSurprisal}, and \cref{seq:ForwardFreeEnergyBoundDerivation4} applies the inequality in \cref{seq:NonadiabaticPotentialInequality} to introduce the nonadiabatic potential ansatz.
Solving the last inequality for \( \Delta F \) gives the forward free-energy bound in \cref{eq:ForwardFreeEnergyBound} of the main text.

\section{Saturating the variational free-energy bounds}
\label{sec:SaturatedFreeEnergyBoundsDerivation}

To see that the forward free-energy bound in \cref{eq:ForwardFreeEnergyBound} of the main text is saturated upon substituting \( V_{\bm\theta} \rightsquigarrow V \), simply note that the substitution renders the right-hand side equal to that of \cref{eq:ForwardFreeEnergy}, which is an equality in the main text.
As for the backward free-energy bound in \cref{eq:BackwardFreeEnergyBound} of the main text, note that the forward-backward likelihood ratio in \cref{eq:PathRelativeDensityBackward} reduces to unity for almost every trajectory as shown in \cref{sec:PathRelativeDensityDerivation} of this document; consequently, the left-hand side of \cref{eq:PathKullbackLeiblerBackward} is reduced to zero and must be equal to the nonnegative right-hand side.

\bibliography{main}